\documentclass[english,american,aip,floatfix,reprint,onecolumn,sort&compress]{revtex4-1}
\usepackage[T1]{fontenc}
\usepackage[latin9]{inputenc}
\usepackage{geometry}
\usepackage{color}
\usepackage{babel}
\usepackage{float}
\usepackage{algorithm2e}
\usepackage{amsmath}
\usepackage{amssymb}
\usepackage{graphicx}

\makeatletter

\@ifundefined{textcolor}{}
{%
  \definecolor{BLACK}{gray}{0}
  \definecolor{WHITE}{gray}{1}
  \definecolor{RED}{rgb}{1,0,0}
  \definecolor{GREEN}{rgb}{0,1,0}
  \definecolor{BLUE}{rgb}{0,0,1}
  \definecolor{CYAN}{cmyk}{1,0,0,0}
  \definecolor{MAGENTA}{cmyk}{0,1,0,0}
  \definecolor{YELLOW}{cmyk}{0,0,1,0}
}

\usepackage{graphicx}        

\makeatother

\begin{document}
\title{Machine Learning for Molecular Dynamics on Long Timescales}
\author{Frank No\'e}
\address{Freie Universitaet Berlin, Department of Mathematics and Computer Science,
Arnimallee 6, 14195 Berlin}
\email{frank.noe@fu-berlin.de}

\selectlanguage{american}%
\begin{abstract}
Molecular Dynamics (MD) simulation is widely used to analyze the properties
of molecules and materials. Most practical applications, such as comparison
with experimental measurements, designing drug molecules, or optimizing
materials, rely on statistical quantities, which may be prohibitively
expensive to compute from direct long-time MD simulations. Classical
Machine Learning (ML) techniques have already had a profound impact
on the field, especially for learning low-dimensional models of the
long-time dynamics and for devising more efficient sampling schemes
for computing long-time statistics. Novel ML methods have the potential
to revolutionize long-timescale MD and to obtain interpretable models.
ML concepts such as statistical estimator theory, end-to-end learning,
representation learning and active learning are highly interesting
for the MD researcher and will help to develop new solutions to hard
MD problems. With the aim of better connecting the MD and ML research
areas and spawning new research on this interface, we define the learning
problems in long-timescale MD, present successful approaches and outline
some of the unsolved ML problems in this application field.
\end{abstract}
\maketitle

\section{Introduction}

Molecular dynamics (MD) simulation is a widely used method of computational
physics and chemistry to compute properties of molecules and materials.
Examples include to simulate how a drug molecule binds to and inhibits
a protein, or how a battery material conducts ions. Despite its high
computational cost, researchers use MD in order to get a principled
understanding of how the composition and the microscopic structure
of a molecular system translate into such macroscopic properties.
In addition to scientific knowledge, this understanding can be used
for designing molecular systems with better properties, such as drug
molecules or enhanced materials.

MD has many practical problems, but at least four of them can be considered
to be fundamental, in the sense that none of them is trivial for a
practically relevant MD simulation, and there is extensive research
on all of them. We refer to these four fundamental MD problems as
SAME (Sampling, Analysis, Model, Experiment):
\begin{enumerate}
\item \textbf{S}ampling: To compute expectation values via MD simulations
the simulation time needs to significantly exceed the slowest equilibration
process in the molecular system. For most nontrivial molecules and
materials, the presence of rare events and the sheer cost per MD time
step make sufficient direct sampling unfeasible.
\item \textbf{A}nalysis: If enough statistics can be collected, we face
huge amounts of simulation data (e.g., millions of time steps, each
having 100,000s of dimensions). How can we analyze such data and obtain
comprehensive and comprehensible models of the most relevant states,
structures and events sampled in the data?
\item \textbf{M}odel: MD simulations employ an empirical model of the molecular
system studied. As the simulation computes forces from an energy model,
this model is often referred to a MD force field. MD energy models
are build from molecular components fitted to quantum mechanical and
experimental data. The accuracy of such a model is limited by the
accuracy of the data used and the errors involved in transferring
the training data usually obtained for small molecules to the often
larger molecules simulated.
\item \textbf{E}xperiment: Experiments and simulations cannot access the
same observables. While in MD simulation, the positions and velocities
of all particles are available at all times, experiments usually probe
complex functions of the positions and velocities, such as emission
or absorption spectra of certain types of radiation. Computing these
functions from first principles often requires the solution of a quantum-mechanical
calculation with an accuracy that is unfeasible for a large molecular
system. The last problem thus consists of finding good approximations
to compute how an experiment would ``see'' a given MD state.
\end{enumerate}
Machine Learning (ML) has the potential to tackle these problems,
and has already had profound impact on alleviating them. Here I will
focus on the analysis problem and its direct connections to the sampling
problem specifically for the case of long-time MD where these problems
are most difficult and interesting. I believe that the solution of
these problems lies on the interface between Chemical Physics and
ML, and will therefore describe these problems in a language that
should be understandable to audiences from both fields.

Let me briefly link to MD problems and associated ML approaches not
covered by this chapter. The present description focuses on low-dimensional
models of long-time MD and these can directly be employed to attack
the sampling problem. The direct effect of these models is that short
MD simulations that are individually not sampling all metastable states
can be integrated, and thus an effective sampling that is much longer
than the individual trajectory length, and on the other of the total
simulation time can be reached \citep{PrinzEtAl_JCP10_MSM1}. The
sampling efficiency can be further improved by adaptively selecting
the starting points of MD simulations based on the long-time MD model,
and iterating this process \citep{hinrichs-pande:jcp:2007:adaptive-sampling,WojtasRouxBerneche_JCTC13_AdaptiveUmbrellaSampling,PretoClementi_PCCP14_AdaptiveSampling,DoerrDeFabritiis_JCTC14_OnTheFly,ZimmermanBowman_JCTC15_FAST,DoerrEtAl_JCTC16_HTMD,PlattnerEtAl_NatChem17_BarBar}.
This approach is called ``adaptive sampling'' in the MD community,
which is an active learning approach in ML language. Using this approach,
time-scales beyond seconds have been reached and protein-protein association
and dissociation has recently been sampled for the first time with
atomistic resolution \citep{PlattnerEtAl_NatChem17_BarBar}. 

A well establish approach to speed up rare events in MD is to employe
so-called enhanced sampling methods that change the thermodynamic
conditions (temperature, adding bias potentials, etc.) \citep{Torrie_JCompPhys23_187,Grubmueller_PhysRevE52_2893,Hansmann1997,LaioParrinello_PNAS99_12562,FukunishiEtAl_JCP02_HamiltonianREMD},
and to subsequently reweight to the unbiased target ensemble \citep{FerrenbergSwendsen_PRL89_WHAM,BartelsKarplus_JCC97_WHAM-ML,Bartels_CPL00_MBAR,GallicchioLevy_JPCB05_TWHAM,ShirtsChodera_JCP08_MBAR,MeyWuNoe_xTRAM}.
Recently, ML methods have been used to adaptively learn optimal biasing
functions in such approaches \citep{ValssonParrinello_PRL14_VariationalMeta,RibeiroTiwary_JCP18_RAVE}.
A conceptually different approach to sampling is the Boltzmann Generator
\citep{NoeWu_18_BoltzmannGenerators}, a directed generative network
to directly draw statistically independent samples from equilibrium
distributions. While these approaches are usually limited to compute
stationary properties, ML-based MD analysis models have recently been
integrated with enhance sampling methods in order to also compute
unbiased dynamical properties \citep{WuMeyRostaNoe_JCP14_dTRAM,WuNoe_MMS14_TRAM1,RostaHummer_DHAM,WuEtAL_PNAS16_TRAM}.
These methods can now also access all-atom protein dynamics beyond
seconds timescales \citep{PaulEtAl_PNAS17_Mdm2PMI}.

ML methods that use MD trajectory data to obtain a low-dimensional
models of the long-time dynamics are extensively discussed here. Not
discussed are manifold learning methods that purely use the data distribution,
such as kernel PCA \citep{SchoelkopfSmolaMueller_NeurComp98_kPCA},
isomap \citep{Tenenbaum_Science290_2319,DasEtAl_PNAS08_Isomap} or
diffusion maps \citep{CoifmanLafon_PNAS05_DiffusionMaps,RohrdanzClementi_JCP134_DiffMaps}.
Likewise, there is extensive research on geometric clustering methods
-- both on the ML and the MD application side -- which only plays
a minor role in the present discussion. 

Learning an accurate MD model -- the so-called force-field problem
-- is one of the basic and most important problems of MD simulation.
While this approach has traditionally been addressed by relatively
\emph{ad hoc} parametrization methods it is now becoming more and
more a well-defined ML problem where universal function approximators
(neural networks or kernel machines) are trained to reproduce quantum-mechanical
potential energy surfaces with high accuracy \citep{BehlerParrinello_PRL07_NeuralNetwork,RuppEtAl_PRL12_QML,BartokKondorCsanyi_PRB13_SOAP,SchuettEtAl_NatComm17_tensornetworksMD,Schuett_SchNet,BereauEtAl_JCP18_Physical}.
See other chapters in this book for more details. A related approach
to scale to the next-higher length-scale is the learning of coarse-grained
MD models from all-atom MD data \citep{WangEtAl_arxiv18_MLCoarseGraining,ZhangEtAl_JCP18_DeePCG,WangBombarelli_Autograin}.
These approaches have demonstrated that they can reach high accuracy,
but employing the kernel machine or neural network to run MD simulations
is still orders of magnitude slower than simulating a highly optimized
MD code with an explicitly coded model. Achieving high accuracy while
approximately matching the computational performance of commonly used
MD codes is an important future aim.

Much less ML work has been done on the interpretation and integration
of experimental data. MD models are typically parametrized by combining
the matching of energies and forces from quantum-mechanical simulations
with the matching of thermodynamic quantities measured by experiments,
such as solvation free energies of small molecules. As yet, there
is no rigorous ML method which learns MD models following this approach.
Several ML methods have been proposed to integrate simulation data
on the level of a model learned from MD simulation data (e.g., a Markov
state model), typically by using information-theoretic principles
such as maximum entropy or maximum caliber \citep{HummerKoefinger_JCP15_Bayesian,OlssonEtAl_PNAS17_AugmentedMarkovModels,DixitDill_JCTRC18_MaxCalMSM}.
Finally, there is an emerging field of ML methods that predict experimental
quantities, such as spectra, from chemical or molecular structures,
which is an essential task that needs to be solved to perform data
integration between simulation and experiment. An important step-stone
for improving our ability to predict experimental properties are the
availability of training datasets where chemical structures, geometric
structures and experimental measurements under well-defined conditions
are linked. 

\section{Learning Problems for Long-time Molecular Dynamics}

\subsection{What would we like to compute?}

The most basic quantitative aim of MD is to compute equilibrium expectations.
When $\mathbf{x}$ is state of a molecular system, such coordinates
and velocities of the atoms in a protein system in a periodic solvent
box, the average value of an observable $A$ is given by: 
\begin{equation}
\mathbb{E}[A]=\int A(\mathbf{x})\,\mu(\mathbf{x})\,\mathrm{d}\mathbf{x}\label{eq:stationary_expectation}
\end{equation}
where $\mu(\mathbf{x})$ is the equilibrium distribution, \emph{i.e.},
the probability to find a molecule in state $\mathbf{x}$ at equilibrium
conditions. A common choice is the Boltzmann distribution in the canonical
ensemble at temperature $T$:
\begin{equation}
\mu(\mathbf{x})\propto\mathrm{e}^{-\frac{U(\mathbf{x})}{k_{B}T}}\label{eq:Boltzmann_distribution}
\end{equation}
where $U(\mathbf{x})$ is a potential energy and the input constant
$k_{B}T$ is the mean thermal energy per degree of freedom. The observable
$A$ can be chosen to compute, \emph{e.g.}, the probability of a protein
to be folded at a certain temperature, or the probability for a protein
and a drug molecule to be bound at a certain drug concentration, which
relates to how much the drug inhibits the protein's activity. Other
equilibrium expectations, such as spectroscopic properties, do not
directly translate to molecular function, but are useful to validate
and calibrate simulation models. 

Molecules are not static but change their state $\mathbf{x}$ over
time. Under equilibrium conditions, these dynamical changes are due
to thermal fluctuations, leading to trajectories that are stochastic.
Given configuration $\mathbf{x}_{t}$ at time $t$, the probability
of finding the molecule in configuration $\mathbf{x}_{t+\tau}$ at
a later time can be expressed by the transition density $p_{\tau}$:
\begin{equation}
\mathbf{x}_{t+\tau}\sim p_{\tau}(\mathbf{x}_{t+\tau}\mid\mathbf{x}_{t}).\label{eq:transition_density}
\end{equation}
Thus, a second class of relevant quantities is that of dynamical expectations:
\begin{equation}
\mathbb{E}[G;\tau]=\int\int\mu(\mathbf{x}_{t})\,p_{\tau}(\mathbf{x}_{t+\tau}\mid\mathbf{x}_{t})\,G(\mathbf{x}_{t},\mathbf{x}_{t+\tau})\,\mathrm{d}\mathbf{x}_{t}\,\mathrm{d}\mathbf{x}_{t+\tau}\label{eq:dynamical_expectation}
\end{equation}
As above, the observable $G$ determines which dynamical property
we are interested in. With an appropriate choice we can measure the
average time a protein takes to fold or unfold, or dynamical spectroscopic
expectations such as fluorescence correlations or dynamical scattering
spectra.

\subsection{What is Molecular Dynamics?}

MD simulation mimics the natural dynamics of molecules by time-propagating
the state of a molecular system, such coordinates and velocities of
the atoms in a protein system in a periodic solvent box. MD is a Markov
process involving deterministic components such as the gradient of
a model potential $U(\mathbf{x})$ and stochastic components, e.g.
from a thermostat. The specific choice of these components determine
the transition density (\ref{eq:transition_density}). Independent
of these choices, a reasonable MD algorithm should be constructed
such that it samples from $\mu(\mathbf{x})$ in the long run: 
\begin{equation}
\lim_{\tau\rightarrow\infty}p_{\tau}(\mathbf{x}_{t+\tau}\mid\mathbf{x}_{t})=\mu(\mathbf{x})\propto\mathrm{e}^{-U(\mathbf{x})/k_{B}T}.\label{eq:convergence_to_stationary}
\end{equation}
Thus, if a long enough MD trajectory can be generated, the expectation
values (\ref{eq:stationary_expectation}) and (\ref{eq:dynamical_expectation})
can be computed as direct averages. Unfortunately, this idea can only
be implemented directly for very small and simple molecular systems.
Most of the interesting molecular systems involve rare events, and
as a result generating MD trajectories that are long enough to compute
the expectation values (\ref{eq:stationary_expectation}) and (\ref{eq:dynamical_expectation})
by direct averaging becomes unfeasible. For example, the currently
fastest special-purpose supercomputer for MD, Anton II, can generate
simulations on the order of 50 $\mu$s per day for a protein system
\citep{ShawEtAl_Anton2}. The time for two strongly binding proteins
to spontaneously dissociate can take over an hour, corresponding to
a simulation time of a century for single event \citep{PlattnerEtAl_NatChem17_BarBar}.

\subsection{Learning Problems for long-time MD}

Repeated sampling from $p_{\tau}(\mathbf{x}_{t+\tau}\mid\mathbf{x}_{t})$
``simulates'' the MD system in time steps of length $\tau$ and
will, due to (\ref{eq:convergence_to_stationary}), result in configurations
sampled from $\mu(\mathbf{x}_{t})$. Hence, knowing $p_{\tau}(\mathbf{x}_{t+\tau}\mid\mathbf{x}_{t})$
is sufficient to compute any stationary or dynamical expectation (\ref{eq:stationary_expectation},\ref{eq:dynamical_expectation}).
The primary ML problem for long-time MD is thus to learn a model of
the probability distribution $p_{\tau}(\mathbf{x}_{t+\tau}\mid\mathbf{x}_{t})$
from simulation data pairs $(\mathbf{x}_{t},\mathbf{x}_{t+\tau})$
which allows $\mathbf{x}_{t+\tau}\sim p_{\tau}(\mathbf{x}_{t+\tau}\mid\mathbf{x}_{t})$
to be efficiently sampled. However, this problem is almost never addressed
directly, because it is unnecessarily difficult. Configurations $\mathbf{x}$
live in a very high-dimensional space (typically $10^{3}$ to $10^{6}$
dimensions), the probability distributions $p_{\tau}(\mathbf{x}_{t+\tau}\mid\mathbf{x}_{t})$
and $\mu(\mathbf{x})$ are multimodal and complex such that direct
sampling is not tractable, and because of the exponential relationship
between energies and probabilities (\ref{eq:Boltzmann_distribution}),
small mistakes in sampling $\mathbf{x}$ will lead to completely unrealistic
molecular structures. 

Because of these difficulties, ML methods for long-time MD usually
take the detour of finding a low-dimensional \emph{representation},
often called latent space representation, $\mathbf{y}=E(\mathbf{x})$,
using the encoder $E$, and learning the dynamics in that space
\[
\begin{array}{ccc}
\text{\ensuremath{\mathbf{x}}}_{t} & \overset{E}{\longrightarrow} & \text{\ensuremath{\mathbf{y}}}_{t}\\
\mathrm{MD}\downarrow\:\:\:\:\:\:\:\:\:\:\: &  & \:\:\:\:\:\downarrow\mathbf{P}\\
\text{\ensuremath{\mathbf{x}}}_{t+\tau} & \overset{D/G}{\longleftarrow} & \text{\ensuremath{\mathbf{y}}}_{t+\tau}
\end{array}
\]
A relatively recent but fundamental insight is that for many MD systems
there exists a natural low-dimensional representation in which the
stationary and dynamical properties can be represented exactly if
we give up time resolution by choosing a large lag time $\tau$. Thus,
for long-time MD the intractable problem to learn $p_{\tau}(\mathbf{x}_{t+\tau}\mid\mathbf{x}_{t})$
can be broken down into three learning problems (LPs) out of which
two are much less difficult, and the third one does not need to be
solved in order to compute stationary or dynamical expectations (\ref{eq:stationary_expectation},\ref{eq:dynamical_expectation}),
and that will be treated in the remainder of the article:
\begin{enumerate}
\item \textbf{LP1: Learn propagator $\mathbf{P}$ in representation $\mathbf{y}$}.
The simplest problem is to learn a model to propagate the latent state
$\mathbf{y}_{t}$ in time for a given encoding $E(\mathbf{x}_{t})$.
This model is often linear using the propagator matrix $\mathbf{P}$,
and hence shallow learning methods such as regression are used. In
addition to obtaining an accurate model, it is desirable for $\mathbf{P}$
to be compact and easily interpretable/readable for a human specialist.
\item \textbf{LP2: Learn encoding $E$ to representation $\mathbf{y}$}.
Learning the generally nonlinear encoding $\mathbf{y}=E(\mathbf{x})$
is a harder problem. Both shallow methods (Regression in kernel and
feature spaces, clustering and likelihood maximization) as well as
deep methods (neural networks) are used. LP1 and LP2 can be coupled
to an end-to-end learning problem for $p_{\tau}\left(E(\mathbf{x}_{t+\tau})\mid E(\mathbf{x}_{t})\right)$.
LP2 has only become a well-defined ML problem recently with the introduction
of a variational approach that defines a meaning loss function for
LP2.
\item \textbf{LP3: Learn decoding $D$/$G$ to configuration space}. The
most difficult problem is to decode the latent representation $\mathbf{y}$
back to configuration space. Because configuration space is much higher
dimensional than latent space, this is an inverse problem. The most
faithful solution is to learn a generator $G$, representing a conditional
probability distribution, $\mathbf{x}\sim G(\mathbf{y})$. This problem
contains the hardest parts of the full learning problem for $p_{\tau}(\mathbf{x}_{t+\tau}\mid\mathbf{x}_{t})$
and addressing it is still in its infancy.
\end{enumerate}

\begin{figure}
\begin{centering}
\includegraphics[width=0.55\columnwidth]{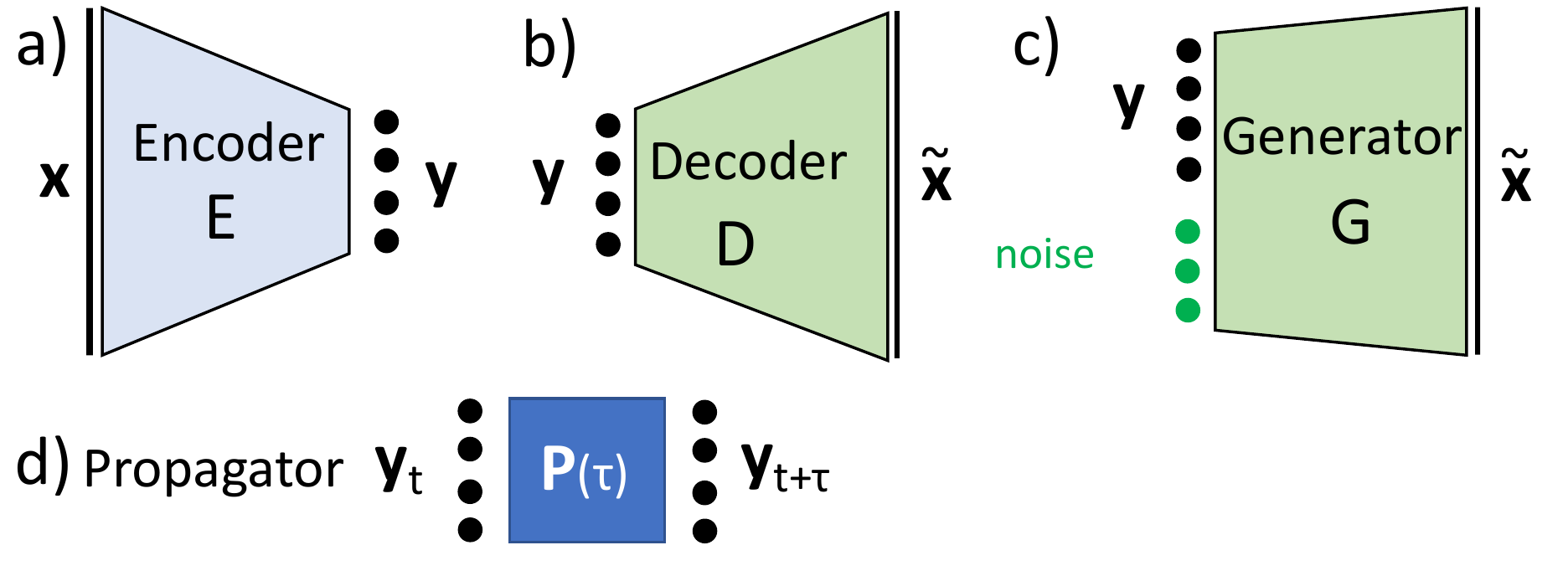}
\par\end{centering}
\caption{\label{fig:building_blocks}Overview of network structures for learning
Markovian dynamical models}
\end{figure}

\paragraph{These learning problems lead to different building blocks that can
be implemented by neural networks or linear methods and can be combined
towards different architectures (Fig. \ref{fig:building_blocks}).}

\section{LP1: Learning Propagator in Feature Space}

\label{sec:LP1}

The simplest and most established learning problem is to learn a propagator,
$\mathbf{P}$, for a given, fixed encoding $E$. Therefore we discuss
this learning problem first before defining what a ``good'' encoding
$E$ is and how to find it. As will be discussed below, for most MD
systems of interest, there exists an encoding $E$ to a \emph{spectral
representation} in which the dynamics is linear and low-dimensional.
Although this spectral representation can often not be found exactly,
it can usually be well enough approximated such that a linear dynamic
model
\begin{equation}
\mathbb{E}\left[\mathbf{y}_{t+\tau}\right]=\mathbf{P}^{\top}\mathbb{E}\left[\mathbf{y}_{t}\right]\label{eq:y-Propagator}
\end{equation}
is an excellent approximation as well. $\mathbb{E}$ denotes an expectation
value over time that accounts for stochasticity in the dynamics, and
can be omitted for deterministic dynamical systems. For example, if
$\mathbf{y}_{t}$ indicates which state the system is in at time $t$,
$\mathbb{E}\left[\mathbf{y}_{t}\right]$ corresponds to a probability
distribution over states. 

Finding a linear model $\mathbf{P}$ is a shallow, unsupervised learning
problem that in many cases has an algebraic expression for the optimum.
Having a linear propagator also has great advantages for the analysis
of the dynamical system. The analyses that can be done depend on the
type of the representation and the mathematical properties of $\mathbf{P}$.
If $E$ performs a one-hot-encoding that indicates which ``state''
the system is in, then the pair $(E,\mathbf{P})$ is called Markov
state model (MSM \citep{SchuetteFischerHuisingaDeuflhard_JCompPhys151_146,SwopePiteraSuits_JPCB108_6571,NoeHorenkeSchutteSmith_JCP07_Metastability,ChoderaEtAl_JCP07,BucheteHummer_JPCB08,PrinzEtAl_JCP10_MSM1}),
and $\mathbf{P}$ is the transition matrix of a Markov chain whose
elements $p_{ij}$ are nonnegative and can be interpreted as the conditional
probabilities to be in a state $j$ at time $t+\tau$ given that the
system was in a state $i$ at time $t$ (Sec. \ref{subsec:LP1_MLMSM}
and \ref{subsec:LP1_MLMSMrev}). For MSMs, the whole arsenal of Markov
chains analysis algorithms is available, e.g. for computing limiting
distributions, first passage times or the statistics of transition
pathways \citep{MetznerSchuetteVandenEijnden_TPT,NoeSchuetteReichWeikl_PNAS09_TPT}.
If the transition matrix additional has a real-valued spectrum, which
is associated with dynamics at thermodynamic equilibrium conditions
(Sec. \ref{subsec:LP1_MLMSMrev}), additional analyses are applicable,
such as the computation of metastable (long-lived) sets of states
by spectral clustering \citep{SchuetteFischerHuisingaDeuflhard_JCompPhys151_146,DeuflhardWeber_LAA05_PCCA+,NoeHorenkeSchutteSmith_JCP07_Metastability}. 

A broader class of propagators arise from encodings $E$ that are
partitions of unity, i.e. where $y_{i}(\mathbf{x})>0$ and $\sum_{i}y_{i}(\mathbf{x})=1$
for all $\mathbf{x}$ \citep{KubeWeber_JCP07_CoarseGraining,MardtEtAl_VAMPnets}.
Such encodings correspond to a ``soft clustering'', where every
configuration $\mathbf{x}$ can still be assigned to a state, but
the assignment is no longer unique. The resulting propagators $\mathbf{P}$
are typically no longer transition matrices whose elements can be
guaranteed to be nonnegative, but they can still be used to propagate
probability densities by means of Eq. (\ref{eq:y-Propagator}), and
if they have a unique eigenvalue of $1$, the corresponding eigenvector
$\boldsymbol{\pi}=[\pi_{i}]$ still corresponds to the unique equilibrium
distribution:
\begin{equation}
\boldsymbol{\pi}=\mathbf{P}^{\top}\boldsymbol{\pi}.\label{eq:stationary_vector}
\end{equation}
For arbitrary functions $E$, we can still use $\mathbf{P}$ to propagate
state vectors according to Eq. (\ref{eq:y-Propagator}), although
these state vectors do no longer have a probabilistic interpretation,
but are simply coefficients that model the configuration in the representation's
basis. Owing to the Markovianity of the model, we can test how well
the time-propagation of the model in time coincides with an estimation
of the model at longer times, by means of the Chapman-Kolmogorov equation:
\begin{equation}
\mathbf{P}^{n}(\tau)\approx\mathbf{P}(n\tau)\label{eq:CKtest}
\end{equation}
In order to implement this equation, one has to decide which matrix
norm should be used to compare the left and right hand side. A common
choice is to compare the leading eigenvalues $\lambda_{i}(\tau)$.
As these decay exponentially with time in a Markov process, it is
common to transform them to relaxation rates or timescales by means
of:
\begin{equation}
t_{i}(\tau)=-\frac{\tau}{\log|\lambda_{i}(\tau)|}\label{eq:relaxation_time}
\end{equation}
A consequence of the Chapman-Kolmogorow equality is that these relaxation
timescales are independent of the lag time $\tau$ at which $\mathbf{P}$
is estimated \citep{SwopePiteraSuits_JPCB108_6571}. For real-valued
eigenvalues, $t_{i}$ corresponds to an ordinary relaxation time of
the corresponding dynamical process. If $\mathbf{P}$ has complex-valued
eigenvalues, $t_{i}$ is the decay time of the envelope of an oscillating
process whose oscillation frequency depends on the phase of $\lambda_{i}$.

\subsection{Loss Function and basis statistics}

Given one or many MD simulation trajectories $\{\mathbf{x}_{t}\}$
and apply $E$ in order to map them to the representation $\{\mathbf{y}_{t}\}$
which define the input to LP1. The basic learning problem is the parameter
estimation problem which consists of obtaining the optimal estimator
$\hat{\mathbf{P}}$ as follows: 
\begin{enumerate}
\item Define a loss function $\mathcal{L}(\mathbf{P};\{\mathbf{y}_{t}\})$
\item Obtain the optimal estimator as $\hat{\mathbf{P}}=\arg\min_{\mathbf{P}}\mathcal{L}(\mathbf{P};\{\mathbf{y}_{t}\})$
\end{enumerate}
As most texts about molecular kinetics do not use the concept of a
loss function, I would like to highlight the importance of a loss
(or score) function from a ML point of view. The difference between
fitting a training data set $\{\mathbf{y}_{t}\}$ and ML is that ML
aims at finding the estimator that performs best on an independent
test data set. To this end we need to not only optimize the parameters
(such as the matrix elements of $\mathbf{P}$), but also hyper-parameters
(such as the size of $\mathbf{P}$), which requires the concept of
a loss function. Another important learning problem is to estimate
the uncertainties of the estimator $\hat{\mathbf{P}}$.

To express the loss function and the optimal estimator of linear propagators
$\mathbf{P}$, we do not actually need the full trajectory $\{\mathbf{y}_{t}\}$,
but only certain sufficient statistics that are usually more compact
than $\{\mathbf{y}_{t}\}$ and thus may require less storage space
and lead to faster algorithms. The most prominent statistics are the
empirical means and covariance matrices:
\begin{align}
\boldsymbol{\mu}_{0} & =\frac{1}{T}\sum_{t=1}^{T-\tau}\mathbf{y}_{t}\label{eq:mean_0}\\
\boldsymbol{\mu}_{\tau} & =\frac{1}{T}\sum_{t=1}^{T-\tau}\mathbf{y}_{t+\tau}\label{eq:mean_t}\\
\mathbf{C}_{00} & =\frac{1}{T}\sum_{t=1}^{T-\tau}\mathbf{y}_{t}\mathbf{y}_{t}^{\top}\label{eq:C_00}\\
\mathbf{C}_{0\tau} & =\frac{1}{T}\sum_{t=1}^{T-\tau}\mathbf{y}_{t}\mathbf{y}_{t+\tau}^{\top}\label{eq:C_0t}\\
\mathbf{C}_{\tau\tau} & =\frac{1}{T}\sum_{t=1}^{T-\tau}\mathbf{y}_{t+\tau}\mathbf{y}_{t+\tau}^{\top}\label{eq:C_tt}
\end{align}
A common modification to (\ref{eq:C_00},\ref{eq:C_tt}) is the so-called
shrinkage estimator that is used in ridge or Tikhonov regularization
\citep{SchaeferStrimmer_05_Shrinkage}. Since many algorithms involve
the inversion of (\ref{eq:C_00},\ref{eq:C_tt}) which might be rank-deficient,
these estimators are often modified by adding a second matrix which
ensures full rank, e.g.:
\begin{align}
\tilde{\mathbf{C}}_{00} & =\mathbf{C}_{00}+\lambda\mathbf{I}\label{eq:C00_Ridge}\\
\tilde{\mathbf{C}}_{\tau\tau} & =\mathbf{C}_{\tau\tau}+\lambda\mathbf{I}\label{eq:Ctt_Ridge}
\end{align}
where the small number $\lambda$ is a regularization hyper-parameter.

\subsection{Maximum Likelihood and Markov State Models}

\label{subsec:LP1_MLMSM}

The concepts of maximum likelihood estimators and Markov State Models
(MSMs) are naturally obtained by defining the following encoding:
\begin{equation}
y_{t,i}=\begin{cases}
1 & \mathbf{x}_{t}\in S_{i}\\
0 & \mathrm{else}.
\end{cases}\label{eq:one-hot}
\end{equation}
where $S_{1},...,S_{n}$ is a partition of configuration space into
$n$ discrete states, i.e. each point $\mathbf{x}$ is assigned to
exactly one state $S_{i}$, indicated by the position of the $1$
in the encoding vector. In ML, (\ref{eq:one-hot}) is called one-hot
encoding. A consequence of (\ref{eq:one-hot}) is that the covariance
matrix (\ref{eq:C_0t}) becomes:
\[
c_{0\tau,ij}=N_{ij}
\]
where $N_{ij}$ counts the total number of transitions observed from
$i$ to $j$. The covariance matrix (\ref{eq:C_00}) is a diagonal
matrix with diagonal elements
\[
c_{00,ii}=N_{i}=\sum_{j}N_{ij}
\]
where we use $N_{i}$ to count the total number of transitions starting
in state $i$. With this encoding, a natural definition for the propagator
$\mathbf{P}$ is a transition matrix whose elements indicate the transition
probability from any state $i$ to any state $j$ in a time step $\tau$:
\[
p_{ij}=\mathbb{P}\left[y_{t+\tau,j}=1\mid y_{t,i}=1\right]
\]
A natural optimality principle is then the maximum likelihood estimator
(MLE): find the transition matrix $\hat{\mathbf{P}}$ that has the
highest probability to produce the observation $\{\mathbf{y}_{t}\}$.
The likelihood is given by:

\begin{equation}
L\propto\prod_{i,j}p_{ij}^{N_{ij}}.\label{eq:MSM_likelihood}
\end{equation}
Where the last term collects equal transition events along the trajectory
and discards the proportionality factor. Maximizing $L$ is equivalent
to minimizing $-L$. However, as common in likelihood formulations
we instead use $-\log L$ as a loss, which is minimal at the same
$\hat{\mathbf{P}}$ but avoids the product:
\begin{equation}
\mathcal{L}_{\mathrm{ML}}(\mathbf{P};\{\mathbf{y}_{t}\})=-\log L=-\sum_{i,j}N_{ij}\log p_{ij}\label{eq:Ploss_likelihood}
\end{equation}
The MLE $\hat{\mathbf{P}}$ can be easily found by minimizing (\ref{eq:Ploss_likelihood})
with the constraint $\sum_{j}p_{ij}=1$ using the method of Lagrange
multipliers. The result is intuitive: the maximum likelihood transition
probability equals the corresponding fraction of transitions observed
out of each state:
\[
p_{ij}=\frac{N_{ij}}{N_{i}}
\]
In matrix form we can express this estimator as
\begin{equation}
\mathbf{P}=\mathbf{C}_{00}^{-1}\mathbf{C}_{0\tau},\label{eq:ML_MSM}
\end{equation}
an expression that we will find also for other optimization principles.
As we have a likelihood (\ref{eq:MSM_likelihood}), we can also define
priors and construct a full Bayesian estimator that not only provides
the maximum likelihood result (\ref{eq:ML_MSM}), but also posterior
means and variances for estimating uncertainties. Efficient samplers
are known that allow us to sample transition matrices directly from
the distribution (\ref{eq:MSM_likelihood}), and these samples can
be used to compute uncertainties on quantities derived from $\mathbf{P}$
\citep{SinghalPande_JCP123_204909,Singhal_JCP07}.

An important property of a transition matrix is its stationary distribution
$\boldsymbol{\pi}$ (which we will assume to exist and be unique here)
with
\[
\pi_{i}=\int_{\mathbf{x}\in S_{i}}\mu(\mathbf{x})\,\mathrm{d}\mathbf{x}.
\]
$\boldsymbol{\pi}$ that can be computed by solving the eigenvalue
problem (\ref{eq:stationary_vector}).

\subsection{MSMs with Detailed Balance}

\label{subsec:LP1_MLMSMrev}

In thermodynamic equilibrium, i.e., when a molecular system is evolving
purely as a result of thermal energy at a given thermodynamic condition
and no external force is applied, the absolute probability of paths
between any two end-points is symmetric. As a consequence of this,
there exists no cycle in state space which contains net flux in either
direction, and no net work can be extracted from the system, consistently
with the second law of thermodynamics. We call this condition \emph{detailed
balance} and write it as:
\begin{equation}
\mu(\mathbf{x})\,p_{\tau}(\mathbf{y}\mid\mathbf{x})=\mu(\mathbf{y})\,p_{\tau}(\mathbf{x}\mid\mathbf{y})\:\forall\mathbf{x},\mathbf{y},\tau>0\label{eq:detailed_balance_continuous}
\end{equation}
Integrating $\mathbf{x}$ and $\mathbf{y}$ over the sets $S_{i}$
and $S_{j}$ in this equation leads to detailed balance for MSMs:
\begin{equation}
\pi_{i}p_{ij}=\pi_{j}p_{ji}.\label{eq:detailed_balance_discrete}
\end{equation}
When the molecular system is simulated such that equations (\ref{eq:detailed_balance_continuous})
hold, we also want to ensure that the estimator $\hat{\mathbf{P}}$
fulfills the constraint (\ref{eq:detailed_balance_discrete}). Enforcing
(\ref{eq:detailed_balance_continuous}) in the estimator reduces the
number of free parameters and thus improves the statistics. More importantly,
propagators that fulfill (\ref{eq:detailed_balance_continuous}) or
(\ref{eq:detailed_balance_discrete}) have a real-valued spectrum
for which additional analyses can be made (see beginning of Sec. \ref{sec:LP1}).

The trivial estimator (\ref{eq:ML_MSM}) does not fulfill (\ref{eq:detailed_balance_discrete}),
unless $N_{ij}$ is, by chance, a symmetric matrix. Maximum likelihood
estimation with (\ref{eq:detailed_balance_discrete}) as a constraint
can be achieved by an iterative algorithm first developed in \citep{Bowman_JCP09_Villin}
and reformulated as in Algorithm \ref{alg:P_MLE_DB} in \citep{TrendelkampSchroerEtAl_InPrep_revMSM}.
Enforcing (\ref{eq:detailed_balance_discrete}) is only meaningful
if there is a unique stationary distribution, which, requires the
transition matrix to define a fully connected graph. For this reason,
graph algorithms are commonly used to find the largest connected set
of states before estimating an MSM with detailed balance \citep{Bowman_JCP09_Villin,PrinzEtAl_JCP10_MSM1,SchererEtAl_JCTC15_EMMA2}.

\begin{algorithm}
\begin{enumerate}
\item Initialize: $\pi_{i}^{(0)}=\frac{\sum_{j=1}^{n}c_{ij}}{\sum_{i,j=1}^{n}c_{ij}}$
\item Iterate until convergence: $\pi_{i}^{(k+1)}=\sum_{j=1}^{n}\frac{c_{ij}+c_{ji}}{c_{i}/\pi_{i}^{(k)}+c_{j}/\pi_{j}^{(k)}}$
\item $p_{ij}=\frac{(c_{ij}+c_{ji})\pi_{j}}{c_{i}\pi_{i}+c_{j}\pi_{i}}$
\end{enumerate}
\caption{\label{alg:P_MLE_DB}\textbf{Detailed balance $\pi_{i}p_{ij}=\pi_{j}p_{ji}$
with unknown $\boldsymbol{\pi}$} \citep{Bowman_JCP09_Villin,TrendelkampSchroerEtAl_InPrep_revMSM}}
\end{algorithm}
When the equilibrium distribution $\boldsymbol{\pi}$ is known \emph{a
priori} or obtained from another estimator as in \citep{WuMeyRostaNoe_JCP14_dTRAM,TrendelkampNoe_PRX15_RareEventKinetics,WuEtAL_PNAS16_TRAM},
the maximum likelihood estimator can be obtained by the iterative
Algorithm \ref{alg:P_MLE_DBgivenpi} developed in \citep{TrendelkampSchroerEtAl_InPrep_revMSM}:

\begin{algorithm}
\begin{enumerate}
\item Initialize Lagrange parameters: $\lambda_{i}^{(0)}=\frac{1}{2}\sum_{j}(c_{ij}+c_{ji})$
\item Iterate until convergence: $\lambda_{i}^{(k+1)}=\sum_{j,c_{ij}+c_{ji}>0}^{n}\frac{(c_{ij}+c_{ji})\lambda_{i}^{(k)}\pi_{j}}{\lambda_{j}^{(k)}\pi_{i}+\lambda_{i}^{(k)}\pi_{j}}$
\item $p_{ij}=\frac{(c_{ij}+c_{ji})\pi_{j}}{\lambda_{i}\pi_{i}+\lambda_{j}\pi_{i}}$
\end{enumerate}
\caption{\label{alg:P_MLE_DBgivenpi}\textbf{Detailed balance $\pi_{i}p_{ij}=\pi_{j}p_{ji}$
with known $\boldsymbol{\pi}$} \citep{TrendelkampSchroerEtAl_InPrep_revMSM}:}
\end{algorithm}
As for MSMs without detailed balance, methods have been developed
to perform a full Bayesian analysis of MSMs with detailed balance.
No method is known to sample independent transition matrices from
the likelihood (\ref{eq:MSM_likelihood}) subject to the detailed
balance constraints (\ref{eq:detailed_balance_discrete}), however
efficient Markov Chain Monte Carlo methods have been developed and
implemented to this end \citep{Noe_JCP08_TSampling,BacalladoChoderaPande_JCP09_ReversibleT,ChoderaNoe_JCP09_MSMstatisticsII,MetznerNoeSchuette_Sampling,TrendelkampNoe_JCP13_EfficientSampler,TrendelkampSchroerEtAl_InPrep_revMSM,SchererEtAl_JCTC15_EMMA2}.

\subsection{Minimal Regression Error}

\label{subsec:Minimal-Regression-Error}

We can understand equation (\ref{eq:y-Propagator}) as a regression
from $\mathbf{y}_{t}$ onto $\mathbf{y}_{t+\tau}$ where $\mathbf{P}$
contains the unknown coefficients. The regression loss is then directly
minimizing the error in Eq. (\ref{eq:y-Propagator}): 
\[
\min\mathbb{E}_{t}\left[\left\Vert \mathbf{y}_{t+\tau}-\mathbf{P}^{\top}\mathbf{y}_{t}\right\Vert ^{2}\right]
\]
and for a given dataset $\{\mathbf{y}_{t}\}$ we can define matrices
$\mathbf{Y}_{0}=\left(\mathbf{y}_{0},...,\mathbf{y}_{T-\tau}\right)^{\top}$
and $\mathbf{Y}_{\tau}=\left(\mathbf{y}_{\tau},...,\mathbf{y}_{T}\right)^{\top}$
resulting in the loss function:
\begin{equation}
\mathcal{L}_{\mathrm{LSQ}}(\mathbf{P};\{\mathbf{y}_{t}\})=\left\Vert \mathbf{Y}_{0}-\mathbf{Y}_{\tau}\mathbf{P}\right\Vert _{F}^{2}\label{eq:Ploss_least_squares}
\end{equation}
where $F$ indicates the Frobenius norm, i.e. the sum over all squares.
The direct solution of the least squares regression problem in (\ref{eq:Ploss_least_squares})
is identical with the trivial MSM estimator (\ref{eq:ML_MSM}). Thus,
the estimator (\ref{eq:ML_MSM}) is more general than for MSMs --
it can be applied for to any representation $\mathbf{y}_{t}$. Dynamic
Mode Decomposition (DMD) \citep{SchmidSesterhenn_APS08_DMD,RowleyEtAl_JFM09_DMDSpectral,Schmid_JFM10_DMD,TuEtAl_JCD14_ExactDMD}
and Extended Dynamic Mode Decomposition (EDMD) \citep{WilliamsKevrekidisRowley_JNS15_EDMD}
are also using the minimal regression error, although the usually
consider low-rank approximations of $\mathbf{P}$.

In general, the individual dimensions of the encoding $E$ may not
be orthogonal, and if not, the matrix $\mathbf{C}_{00}$ is not diagonal,
but contains off-diagonal elements quantifying the correlation between
different dimensions. When there is too much correlation between them,
$\mathbf{C}_{00}$ may have some vanishing eigenvalues, i.e. not full
rank, causing it not to be invertible or only invertible with large
numerical errors. A standard approach approach in least squares regression
is to then apply the Ridge regularization (Eq. \ref{eq:C00_Ridge}).
Using (\ref{eq:C00_Ridge}) in the estimator (\ref{eq:ML_MSM}) is
called Ridge regression.

\subsection{Variational Approach for Dynamics with Detailed Balance (VAC)}

\label{subsec:VAC}

Instead of using an optimality principle to estimate $\mathbf{P}$
directly, we will now derive a variational principle for the eigenvalues
and eigenvectors of $\mathbf{P}$, from which we can then easily assemble
$\mathbf{P}$ itself. At first, this approach seems to be a complication
compared to the likelihood or least squares approach, but this approach
is key in making progress on LP2 because the variational principle
for $\mathbf{P}$ has a fundamental relation to the spectral properties
of the transition dynamics in configuration space (\ref{eq:transition_density}).
It also turns out that the variational approach leads to a natural
representation of configurations that we can optimize in end-to-end
learning frameworks. We first define the balanced propagator:
\begin{equation}
\tilde{\mathbf{P}}=\mathbf{C}_{00}^{-\frac{1}{2}}\mathbf{C}_{0\tau}\mathbf{C}_{\tau\tau}^{-\frac{1}{2}}.\label{eq:P_balanced}
\end{equation}
In this section, we will assume that detailed balance holds with the
a unique stationary distribution, Eq (\ref{eq:detailed_balance_continuous}).
In the statistical limit this means that $\mathbf{C}_{00}=\mathbf{C}_{\tau\tau}$
holds and $\mathbf{C}_{0\tau}$ is a symmetric matrix. Using these
constraints, we find the stationary balanced propagator:
\begin{equation}
\tilde{\mathbf{P}}=\mathbf{C}_{00}^{-\frac{1}{2}}\mathbf{C}_{01}\mathbf{C}_{00}^{-\frac{1}{2}}=\mathbf{C}_{00}^{\frac{1}{2}}\mathbf{P}\mathbf{C}_{00}^{-\frac{1}{2}}\label{eq:P_balanced_stationary}
\end{equation}

Where we have used Eq. (\ref{eq:y-Propagator}). Due to the symmetry
of $\mathbf{C}_{0\tau}$, $\tilde{\mathbf{P}}$ is also symmetric
and we have the symmetric eigenvalue decomposition (EVD):
\begin{equation}
\tilde{\mathbf{P}}=\tilde{\mathbf{U}}\boldsymbol{\Lambda}\tilde{\mathbf{U}}^{\top}\label{eq:EVD_P_balanced}
\end{equation}
with eigenvector matrix $\tilde{\mathbf{U}}=\left[\tilde{\mathbf{u}}_{1},...,\tilde{\mathbf{u}}_{n}\right]$
and eigenvalue matrix $\boldsymbol{\Lambda}=\mathrm{diag}(\lambda_{1},...,\lambda_{n})$
ordered as $\lambda_{1}\ge\lambda_{2}\ge\cdots\ge\lambda_{n}$. This
EVD is related to the EVD of $\mathbf{P}$ via a basis transformation:
\begin{equation}
\mathbf{P}=\mathbf{C}_{00}^{-\frac{1}{2}}\tilde{\mathbf{U}}\boldsymbol{\Lambda}\left(\tilde{\mathbf{U}}\mathbf{C}_{00}^{-\frac{1}{2}}\right)^{\top}=\mathbf{U}\boldsymbol{\Lambda}\mathbf{U}^{-1}\label{eq:EVD_P}
\end{equation}
such that $\mathbf{U}=\mathbf{C}_{00}^{-\frac{1}{2}}\tilde{\mathbf{U}}$
are the eigenvectors of $\mathbf{P}$, their inverse is given by $\mathbf{U}^{-1}=\mathbf{C}_{00}^{-\frac{1}{2}}\tilde{\mathbf{U}}^{\top}$,
and both propagators share the same eigenvalues. The above construction
is simply a change of viewpoint: instead of optimizing the propagator
$\mathbf{P}$, we might as well optimize its eigenvalues and eigenvectors,
and then assemble $\mathbf{P}$ via Eq. (\ref{eq:EVD_P}).

Now we seek an optimality principle for eigenvectors and eigenvalues.
For symmetric eigenvalue problems such as (\ref{eq:EVD_P_balanced}),
we have the following variational principle: The dominant $k$ eigenfunctions
$\tilde{\mathbf{r}}_{1},...,\tilde{\mathbf{r}}_{k}$ are the solution
of the maximization problem:
\begin{align}
\sum_{i=1}^{k}\lambda_{i} & =\max_{\tilde{\mathbf{f}}_{1},...,\tilde{\mathbf{f}}_{k}}\sum_{i=1}^{k}\frac{\tilde{\mathbf{f}}_{i}^{\top}\tilde{\mathbf{P}}\tilde{\mathbf{f}}_{i}}{\left(\tilde{\mathbf{f}}_{i}^{\top}\tilde{\mathbf{f}}_{i}\right)^{\frac{1}{2}}\left(\tilde{\mathbf{f}}_{i}^{\top}\tilde{\mathbf{f}}_{i}\right)^{\frac{1}{2}}}=\max_{\mathbf{f}_{1},...,\mathbf{f}_{k}}\sum_{i=1}^{k}\frac{\mathbf{f}_{i}^{\top}\mathbf{C}_{0\tau}\mathbf{f}_{i}}{\left(\mathbf{f}_{i}^{\top}\mathbf{C}_{00}\mathbf{f}_{i}\right)^{\frac{1}{2}}\left(\mathbf{f}_{i}^{\top}\mathbf{C}_{00}\mathbf{f}_{i}\right)^{\frac{1}{2}}}\nonumber \\
 & =\sum_{i=1}^{k}\frac{\mathbf{u}_{i}^{\top}\mathbf{C}_{0\tau}\mathbf{u}_{i}}{\left(\mathbf{u}_{i}^{\top}\mathbf{C}_{00}\mathbf{u}_{i}\right)^{\frac{1}{2}}\left(\mathbf{u}_{i}^{\top}\mathbf{C}_{00}\mathbf{u}_{i}\right)^{\frac{1}{2}}}=\left(\mathbf{U}^{\top}\mathbf{C}_{00}\mathbf{U}\right)^{-\frac{1}{2}}\mathbf{U}^{\top}\mathbf{C}_{0\tau}\mathbf{U}\left(\mathbf{U}^{\top}\mathbf{C}_{00}\mathbf{U}\right)^{-\frac{1}{2}}\label{eq:variational_principle_matrix}
\end{align}
This means: we vary a set of vectors $\mathbf{f}_{i}=\mathbf{C}_{00}^{-\frac{1}{2}}\tilde{\mathbf{f}}_{i}$,
and when the so-called Rayleigh quotients on the right hand side are
maximized, we have found the eigenvectors. In this limit, the argument
of the Rayleigh quotient equals the sum of eigenvalues. As the argument
above can be made for every value of $k$ starting from $k=1$, we
have found each single eigenvalue and eigenvector at the end of the
procedure (assuming no degeneracy). This variational principle becomes
especially useful for LP2, because using the variational approach
of conformation dynamics (VAC \citep{NoeNueske_MMS13_VariationalApproach,NueskeEtAl_JCTC14_Variational}),
it can also be shown that the eigenvalues of $\mathbf{P}$ are lower
bounds to the true eigenvalues of the Markov dynamics in configurations
$\mathbf{x}$ (Sec. \ref{subsec:variational-principles}). 

Now we notice that this variational principle can also be understood
as a direct correlation function of the data representation. We define
the \emph{spectral representation} as:
\begin{equation}
\mathbf{y}_{t}^{s}=\left(\mathbf{y}_{t}^{\top}\mathbf{u}_{1},...,\mathbf{y}_{t}^{\top}\mathbf{u}_{n}\right)\label{eq:spectral_representation_DB}
\end{equation}
inserting the estimators for $\mathbf{C}_{00}$ and $\mathbf{C}_{0\tau}$
(Eds. \ref{eq:C_00},\ref{eq:C_0t}) into Eq. (\ref{eq:variational_principle_matrix}),
we have:
\[
\sum_{i=1}^{k}\lambda_{i}=\frac{\sum_{t=1}^{T-\tau}\mathbf{y}_{t}^{s}\mathbf{y}_{t+\tau}^{s\top}}{\sum_{t=1}^{T-\tau}\mathbf{y}_{t}^{s}\mathbf{y}_{t}^{s\top}}=\left(\mathbf{C}_{00}^{s}\right)^{-\frac{1}{2}}\mathbf{C}_{0\tau}^{s}\left(\mathbf{C}_{00}^{s}\right)^{-\frac{1}{2}}
\]
where the superscript $s$ denotes the covariance matrices computed
in the spectral representation. 

The same calculation as above can be performed with powers of the
eigenvalues, \emph{e.g.}, $\sum_{i=1}^{k}\lambda_{i}^{2}$. We therefore
get a whole family of VAC-optimization principles, but two choices
are especially interesting: we define the VAC-1 loss, that is equivalent
to the generalized matrix Rayleigh quotient employed in \citep{McGibbonPande_JCP15_CrossValidation},
as:
\begin{align}
\mathcal{L}_{\mathrm{VAC}-1}(\mathbf{U};\{\mathbf{y}_{t}\}) & =-\mathrm{trace}\left[\left(\mathbf{U}^{\top}\mathbf{C}_{00}\mathbf{U}\right)^{-\frac{1}{2}}\mathbf{U}^{\top}\mathbf{C}_{0\tau}\mathbf{U}\left(\mathbf{U}^{\top}\mathbf{C}_{00}\mathbf{U}\right)^{-\frac{1}{2}}\right]\label{eq:VAC1_loss_a}\\
\mathcal{L}_{\mathrm{VAC}-1}(\{\mathbf{y}_{t}^{s}\}) & =-\mathrm{trace}\left[\left(\mathbf{C}_{00}^{s}\right)^{-\frac{1}{2}}\mathbf{C}_{0\tau}^{s}\left(\mathbf{C}_{00}^{s}\right)^{-\frac{1}{2}}\right].\label{eq:VAC1_loss_b}
\end{align}
The VAC-2 loss is the Frobenius norm, i.e. the sum of squared elements
of the matrix:
\begin{align}
\mathcal{L}_{\mathrm{VAC}-2}(\mathbf{U};\{\mathbf{y}_{t}\}) & =-\left\Vert \left(\mathbf{U}^{\top}\mathbf{C}_{00}\mathbf{U}\right)^{-\frac{1}{2}}\mathbf{U}^{\top}\mathbf{C}_{0\tau}\mathbf{U}\left(\mathbf{U}^{\top}\mathbf{C}_{00}\mathbf{U}\right)^{-\frac{1}{2}}\right\Vert _{F}^{2}\label{eq:VAC2_loss_a}\\
\mathcal{L}_{\mathrm{VAC}-2}(\{\mathbf{y}_{t}^{s}\}) & =-\left\Vert \left(\mathbf{C}_{00}^{s}\right)^{-\frac{1}{2}}\mathbf{C}_{0\tau}^{s}\left(\mathbf{C}_{00}^{s}\right)^{-\frac{1}{2}}\right\Vert _{F}^{2}.\label{eq:VAC2_loss_b}
\end{align}
This loss induces a natural spectral embedding where the variance
along each dimension equals the squared eigenvalue and geometric distances
in this space are related to kinetic distances \citep{NoeClementi_JCTC15_KineticMap}.

\subsection{General Variational Approach (VAMP)}

\label{subsec:VAMP}

The variational approach for Markov processes (VAMP) \citep{WuNoe_VAMP}
generalizes the above VAC approach to dynamics that do not obey detailed
balance and may not even have an equilibrium distribution. We use
the balanced propagator (\ref{eq:P_balanced}) that is now no longer
symmetric. Without symmetry we cannot use the variational principle
for eigenvalues, but there is a similar variational principle for
singular values. We therefore use the singular value decomposition
(SVD) of the balanced propagator:
\begin{equation}
\tilde{\mathbf{P}}=\tilde{\mathbf{U}}\boldsymbol{\Sigma}\tilde{\mathbf{V}}^{\top}\label{eq:SVD_P_balanced}
\end{equation}
Again, this SVD is related to the SVD of $\mathbf{P}$ via a basis
transformation:
\begin{equation}
\mathbf{P}=\mathbf{C}_{00}^{-\frac{1}{2}}\tilde{\mathbf{U}}\boldsymbol{\Sigma}\left(\mathbf{C}_{\tau\tau}^{-\frac{1}{2}}\tilde{\mathbf{V}}\right)^{\top}=\mathbf{U}\boldsymbol{\Sigma}\mathbf{V}^{\top}\label{eq:SVD_P}
\end{equation}
with $\mathbf{U}=\mathbf{C}_{00}^{-\frac{1}{2}}\tilde{\mathbf{U}}$
and $\mathbf{V}=\mathbf{C}_{\tau\tau}^{-\frac{1}{2}}\tilde{\mathbf{V}}$.
Using two sets of search vectors $\mathbf{f}_{i}=\mathbf{C}_{00}^{-\frac{1}{2}}\tilde{\mathbf{f}}_{i}$
and $\mathbf{g}_{i}=\mathbf{C}_{\tau\tau}^{-\frac{1}{2}}\tilde{\mathbf{g}}_{i}$,
we can follow the same line of derivation as above and obtain:
\begin{align*}
\sum_{i=1}^{k}\sigma_{i} & =\max_{\tilde{\mathbf{f}}_{1},...,\tilde{\mathbf{f}}_{k},\tilde{\mathbf{g}}_{1},...,\tilde{\mathbf{g}}_{k}}\sum_{i=1}^{k}\frac{\tilde{\mathbf{f}}_{i}^{\top}\tilde{\mathbf{P}}\tilde{\mathbf{g}}_{i}}{\left(\tilde{\mathbf{f}}_{i}^{\top}\tilde{\mathbf{f}}_{i}\right)^{\frac{1}{2}}\left(\tilde{\mathbf{g}}_{i}^{\top}\tilde{\mathbf{g}}_{i}\right)^{\frac{1}{2}}}\\
 & =\left(\mathbf{U}^{\top}\mathbf{C}_{00}\mathbf{U}\right)^{-\frac{1}{2}}\mathbf{U}^{\top}\mathbf{C}_{0\tau}\mathbf{V}\left(\mathbf{V}^{\top}\mathbf{C}_{\tau\tau}\mathbf{V}\right)^{-\frac{1}{2}}
\end{align*}
Now we define again a spectral representation. If we set $\mathbf{C}_{00}=\mathbf{C}_{\tau\tau}$
(equilibrium case) as above, we can define a single spectral representation,
otherwise we need two sets of spectral coordinates:
\begin{align}
\mathbf{y}_{t}^{s,0} & =\left(\mathbf{y}_{t}^{\top}\mathbf{u}_{1},...,\mathbf{y}_{t}^{\top}\mathbf{u}_{n}\right)\label{eq:spectral_representation_noDB1}\\
\mathbf{y}_{t}^{s,\tau} & =\left(\mathbf{y}_{t}^{\top}\mathbf{v}_{1},...,\mathbf{y}_{t}^{\top}\mathbf{v}_{n}\right)\label{eq:spectral_representation_noDB2}
\end{align}
As in the above procedure, we can define a family of VAMP scores,
where the VAMP-1 and VAMP-2 scores are of special interest:
\begin{align}
\mathcal{L}_{\mathrm{VAMP}-1}(\mathbf{U},\mathbf{V};\{\mathbf{y}_{t}\}) & =-\mathrm{trace}\left[\left(\mathbf{U}^{\top}\mathbf{C}_{00}\mathbf{U}\right)^{-\frac{1}{2}}\mathbf{U}^{\top}\mathbf{C}_{0\tau}\mathbf{V}\left(\mathbf{V}^{\top}\mathbf{C}_{\tau\tau}\mathbf{V}\right)^{-\frac{1}{2}}\right]\label{eq:VAMP1_loss_a}\\
\mathcal{L}_{\mathrm{VAMP}-1}(\{\mathbf{y}_{t}^{s,0},\mathbf{y}_{t}^{s,\tau}\}) & =-\mathrm{trace}\left[\left(\mathbf{C}_{00}^{s}\right)^{-\frac{1}{2}}\mathbf{C}_{0\tau}^{s}\left(\mathbf{C}_{\tau\tau}^{s}\right)^{-\frac{1}{2}}\right].\label{eq:VAMP1_loss_b}
\end{align}
The VAMP-2 score is again related to an embedding where geometric
distance corresponds to kinetic distance \citep{PaulEtAl_VAMP_dimred}:
\begin{align}
\mathcal{L}_{\mathrm{VAMP}-2}(\mathbf{U},\mathbf{V};\{\mathbf{y}_{t}\}) & =-\left\Vert \left(\mathbf{U}^{\top}\mathbf{C}_{00}\mathbf{U}\right)^{-\frac{1}{2}}\mathbf{U}^{\top}\mathbf{C}_{0\tau}\mathbf{V}\left(\mathbf{V}^{\top}\mathbf{C}_{\tau\tau}\mathbf{V}\right)^{-\frac{1}{2}}\right\Vert _{F}^{2}\label{eq:VAMP2_loss_a}\\
\mathcal{L}_{\mathrm{VAMP}-2}(\{\mathbf{y}_{t}^{s,0},\mathbf{y}_{t}^{s,\tau}\}) & =-\left\Vert \left(\mathbf{C}_{00}^{s}\right)^{-\frac{1}{2}}\mathbf{C}_{0\tau}^{s}\left(\mathbf{C}_{\tau\tau}^{s}\right)^{-\frac{1}{2}}\right\Vert _{F}^{2}.\label{eq:VAMP2_loss_b}
\end{align}

\section{Spectral Representation and Variational Approach}

Before turning to LP2, we will relate the spectral decompositions
in the VAC and VAMP approaches described above to spectral representations
of the transition density of the underlying Markov dynamics in $\mathbf{x}_{t}$.
These two representations are connected by variational principles.
Exploiting this principle leads to the result that a meaningful and
feasible formulation of the long-time MD learning problem is to seek
a spectral representation of the dynamics. This representation may
be thought of as a set of collective variables (CVs) pertaining to
the long-time MD, or slow CVs \citep{NoeClementi_COSB17_SlowCVs}.

\subsection{Spectral theory}

\label{subsec:spectral-theory}

We can express the transition density (\ref{eq:transition_density})
as the action of the Markov propagator in continuous-space, and by
its spectral decomposition \citep{SarichNoeSchuette_MMS09_MSMerror,WuNoe_VAMP}:
\begin{align}
p(\mathbf{x}_{t+\tau}) & =\int p(\mathbf{x}_{t+\tau}\mid\mathbf{x}_{t};\tau)p(\mathbf{x}_{t})\:\mathrm{d}\mathbf{x}_{t}\label{eq:density_propagation}\\
 & \approx\sum_{k=1}^{n}\sigma_{k}^{*}\langle p(\mathbf{x}_{t})\mid\phi(\mathbf{x}_{t})\rangle\psi(\mathbf{x}_{t+\tau})\label{eq:Pcont_spectral_decomposition}
\end{align}
The spectral decomposition can be read as follows: The evolution of
the probability density can be approximated as the superposition of
basis functions $\psi$. A second set of functions, $\phi$ is required
in order to compute the amplitudes of these functions. 

In general, Eq. (\ref{eq:Pcont_spectral_decomposition}) is a singular
value decomposition with left and right singular functions $\phi_{k},\psi_{k}$
and true singular values $\sigma_{k}^{*}$ \citep{WuNoe_VAMP}. The
approximation then is a low-rank decomposition in which the small
singular values are discarded. For the special case that dynamics
are in equilibrium and satisfy detailed balance (\ref{eq:detailed_balance_continuous}),
Eq. (\ref{eq:Pcont_spectral_decomposition}) is an eigenvalue decomposition
with the choices:
\begin{align*}
\sigma_{k}^{*} & =\lambda_{k}^{*}(\tau)=\mathrm{e}^{-\tau\kappa_{k}}\in\mathbb{R}\\
\phi_{k}(\mathbf{x}) & =\psi_{k}(\mathbf{x})\mu(\mathbf{x}).
\end{align*}
Hence Eq. (\ref{eq:Pcont_spectral_decomposition}) simplifies: we
only need one set of functions, the eigenfunctions $\psi_{k}$. The
true eigenvalues $\lambda_{k}^{*}$ are real-valued and decay exponentially
with the time step $\tau$ (hence Eq. \ref{eq:relaxation_time}).
The characteristic decay rates $\kappa_{k}$ are directly linked to
experimental observables probing the processes associated with the
corresponding eigenfunctions \citep{BucheteHummer_JPCB08,NoeEtAl_PNAS11_Fingerprints}.
The approximation in Eq. (\ref{eq:Pcont_spectral_decomposition})
is due to truncating all terms with decay rates faster than $\kappa_{n}$.
This approximation improves exponentially with increasing $\tau$. 

Spectral theory makes it clear why learning long-time MD via LP1-3
is significantly simpler than trying to model $p(\mathbf{x}_{t+\tau}\mid\mathbf{x}_{t};\tau)$
directly: For long time steps $\tau$, $p(\mathbf{x}_{t+\tau}\mid\mathbf{x}_{t};\tau)$
becomes intrinsically low-dimensional, and it the problem is thus
significantly simplified by learning to approximate the low-dimensional
representation $(\psi_{1},...,\psi_{n})$ for a given $\tau$.

\subsection{Variational principles}

\label{subsec:variational-principles}

The spectral decomposition of the exact dynamics, Eq. (\ref{eq:Pcont_spectral_decomposition}),
is the basis for the usefulness of the variational approaches described
in Sec. \ref{subsec:VAC} and \ref{subsec:VAMP}. The missing connection
is filled by the following two variational principles. The VAC variational
principle \citep{NoeNueske_MMS13_VariationalApproach} is that for
dynamics obeying detailed balance (\ref{eq:detailed_balance_continuous}),
the eigenvalues $\lambda_{k}$ of a propagator matrix $\mathbf{P}$
via any encoding $\mathbf{y}=E(\mathbf{x})$ are, in the statistical
limit, lower bounds of the true $\lambda_{k}^{*}$. The VAMP variational
principle is more general, as it does not require detailed balance
(\ref{eq:detailed_balance_continuous}), and applies to the singular
values: 
\begin{align*}
\lambda{}_{k} & \le\lambda_{k}^{*}\:\:\:(\mathrm{with}\:\mathrm{DB})\\
\sigma_{k} & \le\sigma_{k}^{*}\:\:\:(\mathrm{no}\:\mathrm{DB}).
\end{align*}
Equality is only achieved for $E(\mathbf{x})=\mathrm{span}(\psi_{1},...,\psi_{n})$
when detailed balance holds, and for $E(\mathbf{x})=\mathrm{span}(\psi_{1},...,\psi_{n},\phi_{1},...,\phi_{n})$
when detailed balance does not hold. Specifically, the eigenvectors
or the singular vectors of the propagator then approximate the individual
eigenfunctions or singular functions (assuming no degeneracy):
\begin{align*}
\lambda_{k}=\lambda_{k}^{*}\longrightarrow & \mathbf{u}_{k}^{\top}E(\mathbf{x})=\psi(\mathbf{x})\\
\sigma_{k}=\sigma_{k}^{*}\longrightarrow & \begin{cases}
\mathbf{u}_{k}^{\top}E(\mathbf{x})=\psi(\mathbf{x})\\
\mathbf{v}_{k}^{\top}E(\mathbf{x})=\phi(\mathbf{x}).
\end{cases}
\end{align*}

As direct consequence of the variational principles above, the loss
function associated with a given embedding $E$ is, in the statistical
limit, also an upper bound to the sum of true eigenvalues:
\begin{align*}
\mathcal{L}_{VAC-r} & \ge-\sum_{k=1}^{n}\left(\lambda_{k}^{*}\right)^{r}\\
\mathcal{L}_{VAMP-r} & \ge-\sum_{k=1}^{n}\left(\sigma_{k}^{*}\right)^{r}
\end{align*}
and for the minimum possible loss, $E$ has identified the dominant
eigenspace or singular space. 

\subsection{Spectral representation learning}

We have seen in Sec. \ref{sec:LP1} (LP1) that a propagator $\mathbf{P}$
can be equivalently represented by its eigenspectrum or singular spectrum.
We can thus define a spectral encoding that attempts to directly learn
the encoding to the spectral representation:
\[
\mathbf{y}_{t}^{s}=E^{s}(\mathbf{x}_{t})
\]
with the choices (\ref{eq:spectral_representation_DB}) or (\ref{eq:spectral_representation_noDB1},\ref{eq:spectral_representation_noDB2}),
depending on whether the dynamics obey detailed balance or not. In
these representations, the dynamics are linear. After encoding to
this representation, the eigenvalues or singular values can be directly
estimated from:
\begin{align}
\boldsymbol{\Lambda} & =\left(\mathbf{R}^{\top}\mathbf{C}_{00}^{s}\mathbf{R}\right)^{-1}\mathbf{R}^{\top}\mathbf{C}_{0\tau}^{s}\mathbf{R}\label{eq:Lambda_est}\\
\boldsymbol{\Sigma} & =\left(\mathbf{U}^{\top}\mathbf{C}_{00}^{s}\mathbf{U}\right)^{-\frac{1}{2}}\mathbf{U}^{\top}\mathbf{C}_{0\tau}^{s}\mathbf{V}\left(\mathbf{V}^{\top}\mathbf{C}_{\tau\tau}^{s}\mathbf{V}\right)^{-\frac{1}{2}}\label{eq:Sigma_est}
\end{align}
Based on these results, we can formulate the learning of the spectral
representation, or variants of it, as the key approach to solve LP2.

\section{LP2: Learning Features and Representation}

\label{sec:LP2}

Above we have denoted the full MD system configuration $\mathbf{x}$
and $\mathbf{y}$ the latent-space representation in which linear
propagators are used. We have seen that there is a special representation
$\mathbf{y}^{s}$. In general there may be a whole pipeline of transformations,
e.g.
\[
\mathbf{x}\rightarrow\mathbf{x}^{f}\rightarrow\mathbf{y}\rightarrow\mathbf{y}^{s}
\]
where the first step is a featurization from full configurations $\mathbf{x}$
to features, e.g. the selection of solute coordinates or the transformation
to internal coordinates such as distances or angles. On the latent
space side $\mathbf{y}$ we may have a handcrafted or a learned spectral
representation. Instead of considering these transformations individually,
we may construct a direct end-to-end learning framework that performs
multiple transformation steps.

To simply notation, we commit to the following notation: $\mathbf{x}$
coordinates are the input to the learning algorithm, whether these
are full Cartesian coordinates of the MD system or already transformed
by some featurization. $\mathbf{y}$ are coordinates in the latent
space representations that are the output of LP2, $\mathbf{y}=E(\mathbf{x})$.
We only explicitly distinguish between different stages within configuration
or latent space (e.g. $\mathbf{y}$ vs $\mathbf{y}^{s}$) when this
distinction is explicitly needed.

\subsection{Suitable and unsuitable loss functions}

\label{sec:LP2-loss_functions}

We first ask: what is the correct formulation for LP2? More specifically:
which of the loss functions introduced in LP1 above are compatible
with LP2? Looking at the sequence of learning problems:
\[
\begin{array}{ccccc}
\mathbf{x} & \stackrel{LP2}{\rightarrow} & \mathbf{y} & \stackrel{LP1}{\rightarrow} & \mathbf{P}\end{array}
\]
It is tempting to concatenate them to an end-to-end learning problem
and try to solve it by minimizing any of the three losses defined
for learning of $\mathbf{P}$ in Sec. \ref{sec:LP1}. However, if
we make the encoding $\mathbf{y}=E(\mathbf{x})$ sufficiently flexible,
we find that only one of the loss functions remains as being suitable
for end-to-end learning, while two others must be discarded as they
have trivial and useless minima:

\textbf{Likelihood loss}: The theoretical minimum of the likelihood
loss (\ref{eq:Ploss_likelihood}) is equal to $0$ and is achieved
if all $p_{ij}\equiv1$ for the transitions observed in the dataset.
However, this maximum can be trivially achieved by learning a representation
that assigns all microstates to a single state, e.g. the first state:
\begin{align*}
\arg\max_{E,P}\mathcal{L}_{\mathrm{ML}}(\mathbf{P};\{E(\mathbf{x}_{t})\}) & =\left(\begin{array}{ccc}
E(\mathbf{x}) & \equiv & 1\\
\mathbf{P} & = & \left(\begin{array}{cccc}
1 & 0 & \cdots & 0\\
n/a & \cdots & \cdots & n/a\\
\vdots &  &  & \vdots
\end{array}\right)
\end{array}\right).
\end{align*}
Maximizing the transition matrix likelihood while varying the encoding
$E$ is therefore meaningless.

\textbf{Regression loss}: A similar problem is encountered with the
regression loss. The theoretical minimum of (\ref{eq:Ploss_least_squares})
is equal to $0$ and is achieved when $\mathbf{y}_{t+\tau}\equiv\mathbf{P}^{\top}\mathbf{y}_{t}$
for all $t$. This, can be trivially achieved by learning the uninformative
representation:
\begin{align*}
\arg\max_{E,P}\mathcal{L}_{\mathrm{LSQ}}(\mathbf{P};\{E(\mathbf{x}_{t})\}) & =\left(\begin{array}{ccc}
E(\mathbf{x}) & \equiv & 1\\
\mathbf{P} & = & \mathbf{Id}
\end{array}\right).
\end{align*}
Minimizing the propagator least squares error while varying the encoding
$E$ is therefore meaningless. See also discussion in \citep{OttoRowley_LinearlyRecurrentAutoencoder}.

\textbf{Variational loss}: The variational loss (VAC or VAMP) does
not have trivial minima. The reason is that, according to the variational
principles \citep{NoeNueske_MMS13_VariationalApproach,WuNoe_VAMP},
the variational optimum coincides with the approximation of the dynamical
dynamical components. A trivial encoding such as $E(\mathbf{x})\equiv1$
only identifies a single component and is therefore variationally
suboptimal. The variational loss is thus the only choice amongst the
losses described in LP1 that can be used to learn both $\mathbf{y}$
and $\mathbf{P}$ in an end-to-end fashion. 

\subsection{Feature selection}

We first address the problem of learning the featurization $\mathbf{x}^{f}$.
We can view this problem as a feature selection problem, i.e. we consider
a large potential set of features and ask which of them leads to an
optimal model of the long-time MD. In this view, learning the featurization
is a model selection problem that can be solved by minimizing the
validation loss.

We can solve this problem by employing the variational losses as follows:
We compute the spectral representation $\mathbf{R}$ or $\mathbf{U},\mathbf{V}$
directly from the training set $\mathbf{X}^{\mathrm{train}}=\left(\mathbf{x}_{0}^{f},...,\mathbf{x}_{T}^{f}\right)^{\top}$
and then recompute the covariance matrices in the validation set $\mathbf{X}^{\mathrm{val}}$.
We then compute the following matrices that are diagonal in the training
set but only approximately diagonal in the validation set. The VAC
and VAMP validation scores can then be computed as $\mathcal{L}_{\mathrm{VAC}}(\mathbf{U}^{\mathrm{train}};\{\mathbf{y}_{t}^{\mathrm{test}}\})$
(Eq. \ref{eq:VAC1_loss_a},\ref{eq:VAC2_loss_a}) or $\mathcal{L}_{\mathrm{VAMP}}(\mathbf{U}^{\mathrm{train}};\{\mathbf{y}_{t}^{\mathrm{test}}\})$
(Eq. \ref{eq:VAMP1_loss_a},\ref{eq:VAMP2_loss_a}). In \citep{SchererEtAl_VAMPselection}
we perform VAMP-2 validation in order to select optimal features for
describing protein folding and find that a combination of torsion
backbone angles and $\mathrm{exp}(-d_{ij})$ with $d_{ij}$ being
the minimum distances between amino acids.

\subsection{Blind Source Separation and TICA}

For a given featurization, a widely used linear learning method to
obtain the spectral representation is an algorithm first introduced
in \citep{Molgedey_94} as a method for blind source separation that
later became known as time-lagged independent component analysis (TICA)
method \citep{HyvaerinenKarhunenOja_ICA_Book,PerezEtAl_JCP13_TICA,SchwantesPande_JCTC13_TICA},
sketched in Algorithm \ref{alg:TICA}. In \citep{PerezEtAl_JCP13_TICA},
it was shown that the TICA algorithm directly follows from the minimization
of the VAC variational loss (\ref{eq:VAC1_loss_b},\ref{eq:VAC2_loss_b})
to best approximate the Markov operator eigenfunctions by a linear
combination of a input features. As a consequence, TICA approximate
the eigenvalues and eigenfunctions of Markov operators that obey detailed
balance (\ref{eq:detailed_balance_continuous}), and therefore approximates
the slowest relaxation processes of the dynamics.

Algorithm \ref{alg:TICA} performs a symmetrized estimation of covariance
matrices in order to guarantee that the eigenvalue spectrum is real.
In most early formulations, one usually symmetrizes only $\mathbf{C}_{0\tau}$
while computing $\mathbf{C}_{00}$ by (\ref{eq:C_00}), which is automatically
symmetric. However these formulations might lead to eigenvalues larger
than 1, which do not correspond to any meaningful relaxation timescale
in the present context -- this problem is avoided by the step 1 in
Algorithm \ref{alg:TICA} \citep{WuEtAl_JCP17_VariationalKoopman}.
Note that symmetrization of $\mathbf{C}_{0\tau}$ introduces an estimation
bias if the data is non-stationary, e.g. because short MD trajectories
are used that have not been started from the equilibrium distribution.
To avoid this problem, please refer to Ref. \citep{WuEtAl_JCP17_VariationalKoopman}
which introduces the Koopman reweighting procedure to estimate symmetric
covariance matrices without this bias, although at the price of an
increased estimator variance.

Furthermore, the covariance matrices in step 1 of Algorithm \ref{alg:TICA}
are computed after removing the mean. Removing the mean has the effect
of removing the eigenvalue $1$ and the corresponding stationary eigenvector,
hence all components return by Algorithm \ref{alg:TICA} approximate
dynamical relaxation processes with finite relaxation timescales estimates
according to Eq. (\ref{eq:relaxation_time}).

The TICA propagator can be directly computed as $\bar{\mathbf{P}}=\bar{\mathbf{C}}_{00}^{-1}\bar{\mathbf{C}}_{0\tau}$,
and is a least-squares result in the sense of Sec. \ref{subsec:Minimal-Regression-Error}.
Various extensions of the TICA algorithm were developed: Kernel formulations
of TICA were first presented in machine learning \citep{HarmelingEtAl_NeurComput03_KernelTDSEP}
and later in other fields \citep{SchwantesPande_JCTC15_kTICA,WilliamsEtAl_Arxiv14_KernelEDMD}.
An efficient way to solve TICA for multiple lag times simultaneously
was introduced as TDSEP \citep{ZieheMueller_ICANN98_TDSEP}. Efficient
computation of TICA for very large feature sets can be performed with
a hierarchical decomposition \citep{PerezNoe_JCTC16_hTICA} compressed
sensing approach \citep{Litzinger_JCTC18_CompressedSensing}. TICA
is closely related to the Dynamic Mode Decomposition (DMD) \citep{SchmidSesterhenn_APS08_DMD,RowleyEtAl_JFM09_DMDSpectral,Schmid_JFM10_DMD,TuEtAl_JCD14_ExactDMD}
and the Extended Dynamic Mode Decomposition (EDMD) algorithms \citep{WilliamsKevrekidisRowley_JNS15_EDMD}.
DMD approximates the left eigenvectors (``modes'') instead of the
Markov operator eigenfunctions described here. EDMD is algorithmically
identical to VAC/TICA, but is in practice also used for dynamics that
do not fulfill detailed balance (\ref{eq:detailed_balance_continuous}),
although this leads to complex-valued eigenfunctions.

\begin{algorithm}
\begin{enumerate}
\item Compute symmetrized mean free covariance matrices
\begin{align*}
\bar{\mathbf{C}}_{00} & =\lambda\mathbf{I}+\sum_{t=1}^{T-\tau}(\mathbf{x}_{t}-\boldsymbol{\mu}_{0})(\mathbf{x}_{t}-\boldsymbol{\mu}_{0})^{\top}+(\mathbf{x}_{t+\tau}-\boldsymbol{\mu}_{\tau})(\mathbf{x}_{t+\tau}-\boldsymbol{\mu}_{\tau})^{\top}\\
\bar{\mathbf{C}}_{0\tau} & =\sum_{t=1}^{T-\tau}(\mathbf{x}_{t}-\boldsymbol{\mu}_{0})(\mathbf{x}_{t+\tau}-\boldsymbol{\mu}_{\tau})^{\top}+(\mathbf{x}_{t+\tau}-\boldsymbol{\mu}_{\tau})(\mathbf{x}_{t}-\boldsymbol{\mu}_{0})^{\top}
\end{align*}
with means $\boldsymbol{\mu}_{0},\boldsymbol{\mu}_{\tau}$ defined
analogously as in (\ref{eq:mean_0}-\ref{eq:mean_t}), where $\lambda$
is an optional ridge parameter.
\item Compute the largest $n$ Eigenvalues and Eigenvectors of:
\[
\bar{\mathbf{C}}_{0\tau}\mathbf{u}_{i}=\lambda_{i}\bar{\mathbf{C}}_{00}\mathbf{u}_{i}
\]
\item Project to spectral representation: $\mathbf{y}_{t}=\left(\mathbf{x}_{t}^{\top}\mathbf{u}_{1},...,\mathbf{x}_{t}^{\top}\mathbf{u}_{n}\right)$
for all $t$
\item Return $\{\mathbf{y}_{t}\}$
\end{enumerate}
\caption{\label{alg:TICA}\textbf{TICA}(\{$\mathbf{x}_{t}$\}, $\tau$, $n$).}
\end{algorithm}

\subsection{TCCA / VAMP}

When the dynamics do not satisfy detailed balance (\ref{eq:detailed_balance_continuous}),
e.g., because they are driven by an external force or field, the TICA
algorithm is not meaningful, as it will not even in the limit of infinite
data approximate the true spectral representation. If detailed balance
holds for the dynamical equations, but the data is non-stationary,
i.e. because short simulation trajectories started from a non-equilibrium
distribution are used, the symmetrized covariance estimation in Algorithm
\ref{alg:TICA} introduces a potentially large bias. 

These problems can be avoided by going from TICA to the time-lagged
canonical correlation analysis (TCCA, Algorithm \ref{alg:TCCA}) which
is a direct implementation of the VAMP approach \citep{WuNoe_VAMP},
i.e. it results from minimizing the VAMP variational loss (\ref{eq:VAMP1_loss_b},\ref{eq:VAMP2_loss_b}),
when approximating the Markov operator singular functions with a linear
combination of features. The TCCA algorithm performs a canonical correlation
analysis (CCA) applied to time series. TCCA returns two sets of features
approximating the left and right singular functions of the Markov
operator and that can be interpreted as the optimal spectral representation
to characterize state of the system ``before'' and ``after'' the
transition with time step $\tau$. For non-stationary dynamical systems,
these representations are valid for particular points in time, $t$
and $t+\tau$ \citep{KoltaiEtAl_Computation18_NonrevMSM}.

VAMP/TCCA as a method to obtain a low-dimensional spectral representation
of the long time MD is discussed in detail in \citep{PaulEtAl_VAMP_dimred},
where the algorithm is used to identify low-dimensional embeddings
of driven dynamical systems, such as an ion channel in an external
electrostatic potential.

\begin{algorithm}
\begin{enumerate}
\item Compute covariance matrices $\mathbf{C}_{00}$, $\mathbf{C}_{0\tau}$,
$\mathbf{C}_{\tau\tau}$ from \{$\mathbf{x}_{t}$\}, as in Eqs. (\ref{eq:C_00}-\ref{eq:C_tt})
or Eqs. (\ref{eq:C00_Ridge},\ref{eq:Ctt_Ridge}).
\item Perform the truncated SVD:
\[
\tilde{\mathbf{P}}=\mathbf{C}_{00}^{-\frac{1}{2}}\mathbf{C}_{0\tau}\mathbf{C}_{\tau\tau}^{-\frac{1}{2}}\approx\mathbf{U}'\mathbf{S}\mathbf{V}'^{\top}
\]
where $\tilde{\mathbf{P}}$ is the propagator for the representations
$\mathbf{C}_{00}^{-\frac{1}{2}}\mathbf{x}_{t}$ and $\mathbf{C}_{\tau\tau}^{-\frac{1}{2}}\mathbf{x}_{t+\tau}$,
$\mathbf{S}=\mathrm{diag}(s_{1},...,s_{k})$ is a diagonal matrix
of the first $k$ singular values that approximate the true singular
values $\sigma_{1},...,\sigma_{k}$, and $\mathbf{U}'$ and $\mathbf{V}'$
consist of the $k$ corresponding left and right singular vectors
respectively.
\item Compute $\mathbf{U}=\mathbf{C}_{00}^{-\frac{1}{2}}\mathbf{U}'$, $\mathbf{V}=\mathbf{C}_{\tau\tau}^{-\frac{1}{2}}\mathbf{V}'$
\item Project to spectral representation: $\mathbf{y}_{t}^{0}=\left(\mathbf{x}_{t}^{\top}\mathbf{u}_{1},...,\mathbf{x}_{t}^{\top}\mathbf{u}_{n}\right)$
and $\mathbf{y}_{t}^{\tau}=\left(\mathbf{x}_{t}^{\top}\mathbf{v}_{1},...,\mathbf{x}_{t}^{\top}\mathbf{v}_{n}\right)$
for all $t$
\item Return $\{(\mathbf{y}_{t}^{0},\mathbf{y}_{t}^{\tau})\}$
\end{enumerate}
\caption{\label{alg:TCCA}\textbf{TCCA}(\{$\mathbf{y}_{t}$\}, $\tau$, $n$)}
\end{algorithm}

\subsection{MSMs based on geometric clustering}

For the spectral representations found by TICA and TCCA, a propagator
$\mathbf{P}(\tau)$ can be computed by means of Eq. (\ref{eq:y-Propagator}),
however this propagator is harder to interpret than a MSM propagator
whose elements correspond to transition probabilities between states.
For this reason, TICA, TCCA and other dimension reduction algorithms
are frequently used as a first step towards building an MSM \citep{PerezEtAl_JCP13_TICA,SchwantesPande_JCTC13_TICA,PerezNoe_JCTC16_hTICA}.
Before TICA and TCCA were introduced into the MD field, MSMs were
directly built upon manually constructed features such as distances,
torsions or in other metric spaces that define features only indirectly,
such as the pairwise distance of aligned molecules \citep{kabsch:acta-cryst:1976:rmsd,theobald:acta-cryst:2005:fast-rmsd}
-- see Ref. \citep{HusicPande_JACS18_MSMReview} for an extensive
discussion.

In this approach, the trajectories in feature space, $\{\mathbf{x}_{t}^{f}\}$,
or in the representation $\{\mathbf{y}_{t}\}$, must be further transformed
into a one-hot encoding (\ref{eq:one-hot}) before the MSM can be
estimated via one of the methods described in Sec. \ref{sec:LP1}.
In other words, the configuration space must be divided into $n$
sets that are associated with the $n$ MSM states. Typically, clustering
methods that somehow group simulation data by means of geometric similarity.
When MSMs were build on manually constructed feature spaces, research
on suitable clustering methods was very active \citep{SwopeEtAl_JPCB108_6582,NoeHorenkeSchutteSmith_JCP07_Metastability,ChoderaEtAl_JCP07,AltisStock_JCP07_dPCA,BucheteHummer_JPCB08,YaoEtAl_JCP09_Mapper,Bowman_JCP09_Villin,Keller_JCP2010_Clustering,JainStock_JCTC12_ProbablePathClustering,BowmanPandeNoe_MSMBook,SchererEtAl_JCTC15_EMMA2,SheongEtAl_JCTC15_APM,HusicPande_JCTC17_Ward}.
Since the introduction of TICA and TCCA that identify a spectral representation
that already approximates the leading eigenfunctions, the choice of
the clustering method has become less critical, and simple methods
such as $k$-means$^{++}$ lead to robust results. The final step
towards an easily interpretable MSM is coarse-graining of $\mathbf{P}$
down to a few states \citep{KubeWeber_JCP07_CoarseGraining,YaoHuang_JCP13_Nystrom,FackeldeyWeber_WIAS17_GenPCCA,GerberHorenko_PNAS17_Categorial,HummerSzabo_JPCB15_CoarseGraining,OrioliFaccioli_JCP16_CoarseMSM,NoeEtAl_PMMHMM_JCP13}.

The geometric clustering step introduces a different learning problem
and objective whose relationship to the original problem of approximating
long-term MD is not clear. Therefore, geometric clustering must be
at the moment regarded as a pragmatic approach to construct an MSM
from a given embedding, but this approach departs from the avenue
of a well-defined machine learning problem.

\subsection{VAMPnets}

\label{sec:VAMPnets}

VAMPnets \citep{MardtEtAl_VAMPnets} were introduced to replace the
complicated and error-prone approach of constructing MSMs by (i) searching
for optimal features $\mathbf{x}^{f}$, (ii) combining them to a representation
$\mathbf{y}$, \emph{e.g.}, \emph{via} TICA, (iii) clustering it,
(iv) estimating the transition matrix $\mathbf{P}$, and (v) coarse-graining
it, by a single end-to-end learning approach in which all of these
steps are replaced by a deep neural network. This is possible because
with the VAC and VAMP variational principles, loss functions are available
that are suitable to train the sequence of learning problems 1 and
2 simultaneously. A similar architecture is used by EDMD with dictionary
learning \citep{LiEtAl_Chaos17_EDMD_DL}, which avoids the problem
of the regression error to collapse to trivial encodings $E$ (Sec.
\ref{sec:LP2-loss_functions}) by fixing some features that are not
learnable.

\begin{figure}
\begin{centering}
\includegraphics[width=0.8\columnwidth]{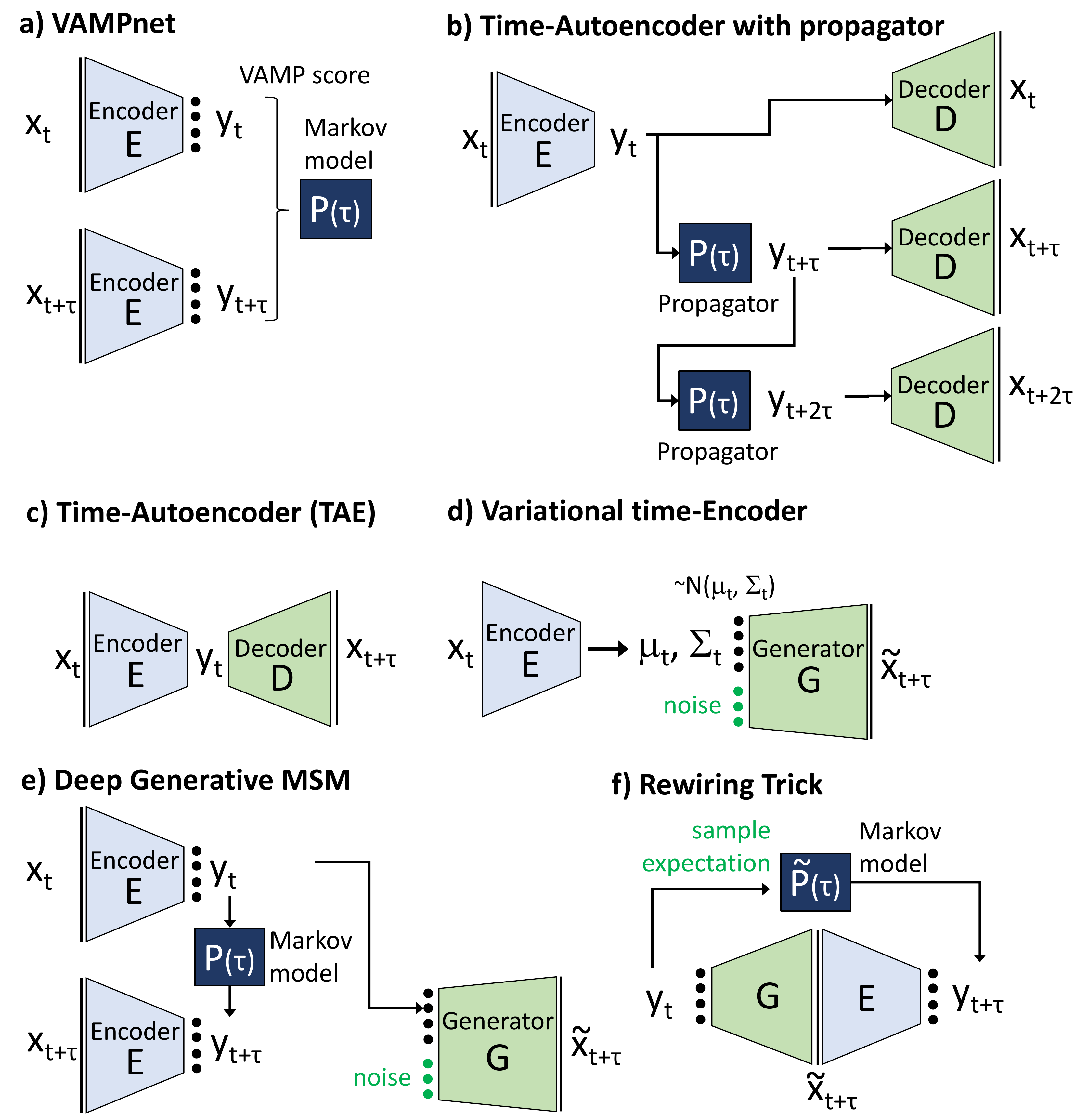}
\par\end{centering}
\caption{\label{fig:network_structures}Overview of network structures for
learning Markovian dynamical models. \textbf{a}) VAMPnets \citep{MardtEtAl_VAMPnets}.
\textbf{b}) Time-autoencoder with propagator \citep{LuschKutzBrunton_DeepKoopman,OttoRowley_LinearlyRecurrentAutoencoder}.
\textbf{c}) time-autoencoder \citep{WehmeyerNoe_TAE}. \textbf{d})
variational time-encoder \citep{Hernandez_PRE18_VTE}. \textbf{e})
Deep Generative Markov State Models \citep{WuEtAl_NIPS18_DeepGenMSM}.
\textbf{f}) The rewiring trick to compute the propagator $\mathbf{P}$
for a deep generative MSM.}
\end{figure}

\begin{figure}
\begin{centering}
\includegraphics[width=1\columnwidth]{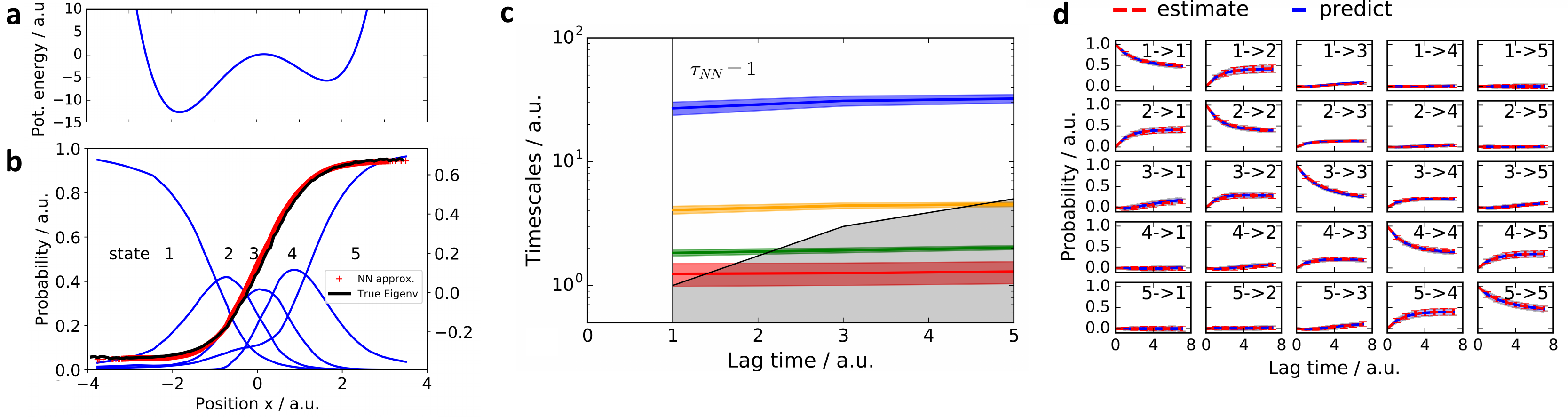}
\par\end{centering}
\centering{}\caption{\label{fig:VAMPnet_1dpot}Figure adapted from \citep{MardtEtAl_VAMPnets}:
Approximation of the slow transition in a bistable potential by a
VAMPnet with one input node ($x$) and five output nodes. (\textbf{a})
Potential energy function $U(x)=x^{4}-6x^{2}+2x$. (\textbf{b}) Eigenvector
of the slowest process calculated \textcolor{black}{by direct numerical
approximation (black) }and approximated by a VAMPnet with five output
nodes (red). Activation of the five Softmax output nodes define the
state membership probabilities (blue). (\textbf{c}) Relaxation timescales
computed from the Koopman model using the VAMPnet transformation.
(\textbf{d}) Chapman-Kolmogorov test comparing long-time predictions
of the Koopman model estimated at $\tau=1$ and estimates at longer
lag times. Panels (c) and (d) report 95\% confidence interval error
bars over 100 training runs.}
 
\end{figure}

VAMPnets contain two network lobes that transform the molecular configurations
found at a time delay $\tau$ along the simulation trajectories (Fig.
\ref{fig:network_structures}a). VAMPnets can be minimized with any
VAC or VAMP variational loss. In Ref. \citep{MardtEtAl_VAMPnets},
the VAMP-2 loss (\ref{eq:VAMP2_loss_b}) was used, which is meaningful
for both dynamics with and without detailed balance. When detailed
balance (\ref{eq:detailed_balance_discrete}) is enforced in the propagator
obtained by (\ref{eq:y-Propagator}), the loss function automatically
becomes VAC-2. VAMPnets may either use two distinct network lobes
to encode the spectral representation of the left and right singular
functions (which is important for non-stationary dynamics \citep{KoltaiCiccottiSchuette_JCP16_MetastabilityNonstationary,KoltaiEtAl_Computation18_NonrevMSM}),
whereas for MD with a stationary distribution we generally use parameter
sharing and have two identical lobes. For dynamics with detailed balance,
the VAMPnet output then encodes the space of the dominant Markov operator
eigenfunctions (Fig. \ref{fig:VAMPnet_1dpot}b).

\begin{figure}
\begin{centering}
\includegraphics[width=1\columnwidth]{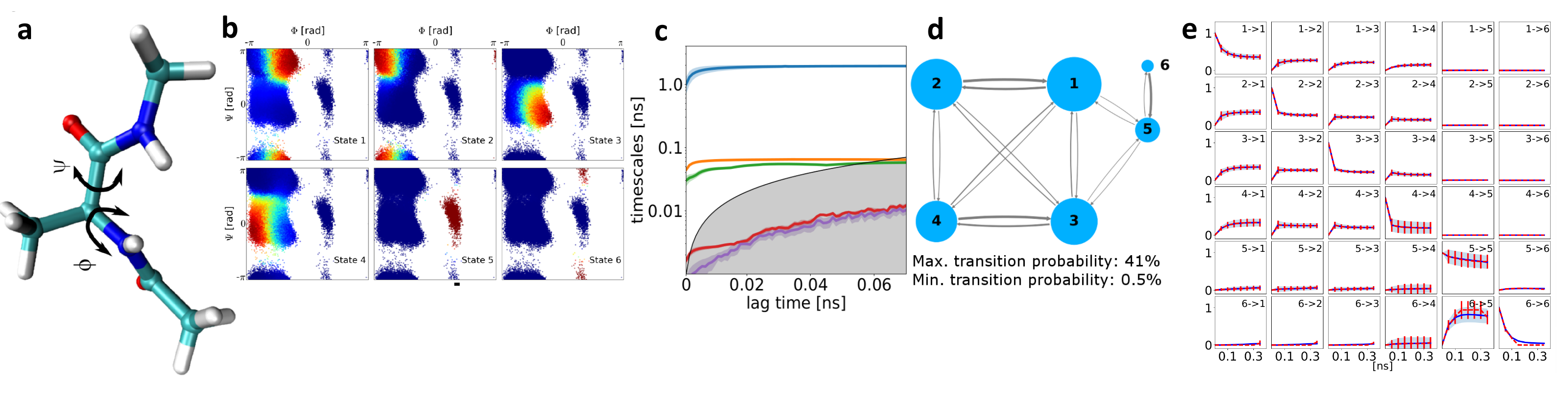}
\par\end{centering}
\centering{}\caption{\label{fig:VAMPnet_ala}Figure adapted from \citep{MardtEtAl_VAMPnets}:
Kinetic model of alanine dipeptide obtained by a VAMPnet with 30 input
nodes ($x,y,z$ Cartesian coordinates of heavy atoms) and six output
nodes. (\textbf{a}) Structure of alanine dipeptide. The main coordinates
describing the slow transitions are the backbone torsion angles $\phi$
and $\psi$, however the neural network inputs are only the Cartesian
coordinates of heavy atoms. (\textbf{b}) Assignment of all simulated
molecular coordinates, plotted as a function of $\phi$ and $\psi$,
to the six Softmax output states. Color corresponds to activation
of the respective output neuron, indicating the membership probability
to the associated metastable state. (\textbf{c}) Relaxation timescales
computed from the Koopman model using the neural network transformation.
(\textbf{d}) Representation of the transition probabilities matrix
of the Koopman model; transitions with a probability lower than 0.5\%
have been omitted. (\textbf{e}) Chapman-Kolmogorov test comparing
long-time predictions of the Koopman model estimated at $\tau=50\,ps$
and estimates at longer lag times. Panels (c) and (e) report 95\%
confidence interval error bars over 100 training runs excluding failed
runs.}
 
\end{figure}

In order to obtain a propagator that can be interpreted as an MSM,
\citep{MardtEtAl_VAMPnets} chose to use a SoftMax layer as an output
layer, thus transforming the spectral representation to a soft indicator
function similar to spectral clustering methods such as PCCA+ \citep{DeuflhardWeber_LAA05_PCCA+,RoeblitzWeber_AdvDataAnalClassif13_PCCA++}.
As a result, the propagator computed by Eq. (\ref{eq:y-Propagator})
is \emph{almost} a transition matrix. It is guaranteed to be a true
transition matrix in the limit where the output layer performs a hard
clustering, i.e. one-hot encoding (\ref{eq:one-hot}). Since this
is not true in general, the VAMPnet propagator may still have negative
elements, but these are usually very close to zero. The propagator
is still valid for transporting probability distributions in time
and can therefore be interpreted as an MSM between metastable states
(Fig. \ref{fig:VAMPnet_ala}d).

The results described in \citep{MardtEtAl_VAMPnets} (see, \emph{e.g.},
Fig. \ref{fig:VAMPnet_1dpot},\ref{fig:VAMPnet_ala}) were competitive
with and sometimes surpassed the state-of-the-art handcrafted MSM
analysis pipeline. Given the rapid improvements of training efficiency
and accuracy of deep neural networks seen in a broad range of disciplines,
we expect end-to-end learning approaches such as VAMPnets to dominate
the field eventually.


\section{LP3 light: Learn Representation and Decoder}

As discussed in Sec. \ref{sec:LP2-loss_functions}, end-to-end learning
combining LP1 and LP2 are limited in their choice of losses applied
to the propagator resulting from LP2: Variational losses can be used,
leading to the methods described in Sec. \ref{sec:LP2}, while using
the likelihood and regression losses are prone to collapse to a trivial
representation that does not resolve the long-time dynamical processes. 

One approach to ``rescue'' these approaches is to add other loss
functions to prevent this collapse to a trivial, uninformative representation
from happening. An obvious choice is to add a decoder that is trained
with some form of reconstruction loss: the representation $\mathbf{r}$
should still contain enough information that the input ($\mathbf{x}$
or $\mathbf{y}$) can be approximately reconstructed. We discuss several
approaches based on this principle. Note that if only finding the
spectral embedding and learning the propagator $\mathbf{P}$ is the
objective, VAMPnets solve this problem directly and employing a reconstruction
loss unnecessarily adds the difficult inverse problem of reconstructing
a high-dimensional variable from a low-dimensional one. However, approximate
reconstruction of inputs may be desired in some applications, and
is the basis for LP3.

\subsection{Time-Autoencoder}

The time-autoencoder \citep{WehmeyerNoe_TAE} shortcuts LP2 and constructs
a direct learning problem between $\mathbf{x}_{t}$ and $\mathbf{x}_{t+\tau}$
(Fig. \ref{fig:network_structures}c).
\begin{equation}
\text{\ensuremath{\mathbf{x}}}_{t}\overset{E}{\longrightarrow}\text{\ensuremath{\mathbf{y}}}_{?}\overset{D}{\longrightarrow}\text{\ensuremath{\mathbf{x}}}_{t+\tau}\label{eq:TAE_scheme}
\end{equation}
The time-autoencoder is trained by reconstruction loss:
\begin{equation}
\mathcal{L}_{\mathrm{TAE}}(E,D;\{\mathbf{x}_{t}\})=\sum_{t=0}^{T-\tau}\left\Vert \mathbf{x}_{t+k\tau}-D\left(E(\mathbf{x}_{t})\right)\right\Vert \label{eq:TAE_loss}
\end{equation}
where $\left\Vert \cdot\right\Vert $ is a suitable norm, \emph{e.g.},
the squared 2-norm.

The TAE has an interesting interpretation: If $E$ and $D$ are linear
transformation, i.e. encoder and decoder matrices $\mathbf{E}\in\mathbb{R}^{N\times n}$,
$\mathbf{D}\in\mathbb{R}^{n\times N}$, the minimum of (\ref{eq:TAE_loss})
is found by VAMP/TCCA, and for data that is in equilibrium and obeys
detailed balance by VAC/TICA \citep{WehmeyerNoe_TAE}. The reverse
interpretation is not true: the solution found by minimizing (\ref{eq:TAE_loss})
does not lead to TICA/TCCA modes, as there is no constraint in the
time-autoencoder for the components $\mathbf{r}_{t}$ -- they only
span the same space. Within this interpretation, the time-autoencoder
can be thought of a nonlinear version of TCCA/TICA in the sense of
being able to find a slow but nonlinear spectral representation. 

Time-autoencoders have several limitations compared to VAMPnets: (1)
Adding the decoder network makes the learning problem more difficult.
(2) As indicated in scheme (\ref{eq:TAE_scheme}), it is not clear
what the time step pertaining to the spectral representation $\mathbf{y}$
is ($t$, $t+\tau$, or something in between), as the time stepping
is done throughout the entire network. (3) Since the decoding problem
from any given $\mathbf{y}$ to $\mathbf{x}_{t+\tau}$ is underdetermined
but the decoder network $D$ is deterministic, it will only be able
to decode to a ``mean'' $\mathbf{x}$ for all $\mathbf{x}$ mapping
to the same $\mathbf{y}$. Thus, time-autoencoders cannot be used
to sample the transition density (\ref{eq:transition_density}) to
generated sequences $\mathbf{x}_{t}\rightarrow\mathbf{x}_{t+\tau}$.

\subsection{Time-Autoencoder with Propagator}

Both \citep{LuschKutzBrunton_DeepKoopman,OttoRowley_LinearlyRecurrentAutoencoder}
have introduced time-autoencoders that additionally learn the propagator
in the spectral representation, and thus fix problem (2) of time-autoencoders,
while problems (1) and (3) still remain.

Instead of scheme (\ref{eq:TAE_scheme}), time-autoencoders with propagator
introduce a time propagation step that makes the time step explicit
for every step:

\begin{equation}
\text{\ensuremath{\mathbf{x}}}_{t}\overset{E}{\longrightarrow}\text{\ensuremath{\mathbf{y}}}_{t}\overset{\mathbf{P}}{\longrightarrow}\text{\ensuremath{\mathbf{y}}}_{t+\tau}\overset{D}{\longrightarrow}\text{\ensuremath{\mathbf{x}}}_{t+\tau}\label{eq:TAEP_scheme}
\end{equation}
where $\mathbf{P}$ is the matrix defined by a $n\times n$ linear
layer. Training this network exclusively with the standard autoencoder
loss would not impose the correct internal structure -- in particular,
it would not be possible to control that $E$ learns only the representation
and $\mathbf{P}$ performs the time step. \citep{LuschKutzBrunton_DeepKoopman,OttoRowley_LinearlyRecurrentAutoencoder}
enforce the dynamical consistency by training several lag times simultaneously
with variants of the following type of loss:
\begin{equation}
\mathcal{L}_{\mathrm{TAE-P}}=\sum_{t=0}^{T-k\tau}\left(\sum_{k=0}^{K}\text{\ensuremath{\alpha}}_{k}\left\Vert \mathbf{x}_{t+k\tau}-D\left(\mathbf{P}^{k}E(\mathbf{x}_{t})\right)\right\Vert +\sum_{k=1}^{K}\text{\ensuremath{\beta}}_{k}\left\Vert E(\mathbf{x}_{t+k\tau})-\mathbf{P}^{k}E(\mathbf{x}_{t})\right\Vert \right)\label{eq:TAEP_loss}
\end{equation}
where $\alpha_{k},\beta_{k}$ are coefficients, the first term correspond
to a autoencoder reconstruction loss and the second term trains the
correct time-propagation of $\mathbf{P}$ in latent space. The number
of lag times, $K$, to be considered is a user-defined choice. Note
that it is not a typical hyper-parameter as matching the dynamics
at more lag times makes the learning problem harder, and thus the
cross-validation score of (\ref{eq:TAEP_loss}) cannot be used to
select $K$. Unrolling the network for $K=2$ results in Fig. \ref{fig:network_structures}b.
This approach works excellently in deterministic (but highly nonlinear)
dynamical systems with short time steps \citep{LuschKutzBrunton_DeepKoopman,OttoRowley_LinearlyRecurrentAutoencoder}. 

In stochastic systems such as MD, it appears more difficult to learn
$\mathbf{r}_{t}$ and $\mathbf{P}$ such that they span the spectral
components of the underlying propagator and recover its largest eigenvalues.
While this observation needs more study, potential explanations are
that in long-time MD we need large time steps $\tau$, in order to
make the spectral representation learning problem low-dimension (see
Sec. \ref{subsec:spectral-theory}), and that the stochastic fluctuations
are large which makes learning a decoder $D$ difficult.

\subsection{Variational (time-)Autoencoders}

\label{subsec:LP3light_VAEs}

Several recent approaches employ variational autoencoders (VAEs) for
the long-time MD or related learning problems. Variational autoencoders
\citep{KingmaWelling_ICLR14_VAE} learn to sample a probability distribution
that approximates the distribution underlying observation data. To
this end, VAEs employ variational Bayesian inference \citep{FoxRoberts_AIRev12_TutorialVariationalBayes}
in order to approximately minimize the KL divergence between the generated
and the observed distribution. VAEs have a similar structure as usual
autoencoders, with an inference network mapping from a high-dimensional
variable $\mathbf{x}$ to a typically lower-dimensional latent variable
$\mathbf{r}$, and attempting to reconstruct $\mathbf{x}$ in a decoder
network. The main difference is that every latent point $\mathbf{r}$
encodes the moments of a distribution which are used to sample $\mathbf{x}$
such that the distributions become similar.

VAEs have been used in RAVE \citep{RibeiroTiwary_JCP18_RAVE} for
enhancing the sampling by identifying a space of ``reaction coordinates''
in which MD sampling can be efficiently driven, and in Autograin \citep{WangBombarelli_Autograin}
to find a way to coarse-grain a molecule into effective beads. Both
methods use VAEs without an inference network that employs a time
step $\tau$, and therefore they address learning problems that are
conceptually different from long-lime MD learning problem as treated
here. 

A much more closely related work are variational time-encoders \citep{Hernandez_PRE18_VTE}
(Fig. \ref{fig:network_structures}d), which employ a VAE between
time steps $\mathbf{x}_{t}$ at the input and $\mathbf{x}_{t+\tau}$
at the output:
\begin{align*}
\mathbf{x}_{t}\overset{E}{\longrightarrow}\mu(\mathbf{x}_{t})\rightarrow\mathbf{y}_{t}\rightarrow & \oplus\rightarrow\mathbf{y}_{t+\tau}\overset{D}{\longrightarrow}\mathbf{x}_{t+\tau}\\
 & \uparrow\\
 & \mathcal{N}(0,1)
\end{align*}
As \citep{Hernandez_PRE18_VTE} note, this approach does not achieve
the sampling of the $\mathbf{x}_{t+\tau}$ distribution (the variational
theory underlying VAEs requires that the same type of variable is
used at input and output) and hence does not act as a propagator $\mathbf{x}_{t}\rightarrow\mathbf{x}_{t+\tau}$,
but succeeds in learning a spectral representation of the system.
For this reason, the variational time-encoder is listed in this section
rather than in LP3. 

\section{LP3 heavy: Learn Generative Models}

The full solution of LP3 involves learning to generate samples $\mathbf{x}_{t+\tau}$
from the lower-dimensional feature embedding or spectral representation.
This is a very important goal as its solution would yield an ability
to sample the MD propagator $\mathbf{x}_{t}\rightarrow\mathbf{x}_{t+\tau}$
at long time-steps $\tau$, which would yield a very efficient simulator.
However, because of the high dimensionality of configuration space
and the complexity of distributions there, this aim is extremely difficult
and still in its infancy. 

Clearly standard tools for learning directed generative networks,
such as Variational Autoencoders \citep{KingmaWelling_ICLR14_VAE}
and generative adversarial nets \citep{GoodfellowEtAl_GANs} are ``usual
suspects'' for the solution of this problem. However, existing applications
of VAEs and GANs on the long-time MD problem have focused on learning
a latent representation that is suitable to encode the long-time processes
or a coarse-graining, and the decoder has been mostly used to regularize
the problem (Sec. \ref{subsec:LP3light_VAEs}). The first approach
to actually reconstruct molecular structures in configuration space,
so as to achieve long-time-step sampling, was made in \citep{WuEtAl_NIPS18_DeepGenMSM},
which will be analyzed in some detail below.

\subsection{Deep Generative MSMs}

The deep generative MSMs described \citep{WuEtAl_NIPS18_DeepGenMSM}
(Fig. \ref{fig:network_structures}e), we propose to address LP1-3
in the following manner. We first formulate a machine learning problem
to learn the following two functions:
\begin{itemize}
\item An probabilistic encoding of the input configuration to a low-dimensional
latent space, $\mathbf{x}_{t}\rightarrow E(\mathbf{x}_{t})$. Similar
to VAMPnets with a probabilistic output (Sec. \ref{sec:VAMPnets}),
$\boldsymbol{\chi}$ has $n$ elements, and each element represents
the probability of configuration $\mathbf{x}$ to be in a metastable
(long-lived) state $i$:
\[
E_{i}(\mathbf{x})=\mathbb{P}(\mathbf{x}_{t}\in\text{state }i\mid\mathbf{x}_{t}=\mathbf{x}).
\]
Consequently, these functions are nonnegative ($E_{i}(x)\ge0\:\:\forall x$)
and sum up to one ($\sum_{i}E_{i}(x)=1\:\:\forall x$). The functions
$E(x)$ can, e.g., be represented by a neural network mapping from
$\mathbb{R}^{d}$ to $\mathbb{R}^{m}$ with a SoftMax output layer. 
\item An $n$-element probability distribution $\mathbf{q}(\mathbf{x};\tau)=\left(q_{1}(\mathbf{x};\tau),...,q_{n}(\mathbf{x};\tau)\right)$,
which assigns to each configuration $\mathbf{x}$ a probability density
that a configuration that was in metastable state $i$ at time $t$,
will ``land'' in $\mathbf{x}$ at time $t+\tau$:
\[
q_{i}(\mathbf{x};\tau)=\mathbb{P}(\mathbf{x}_{t+\tau}=\mathbf{x}|\mathbf{x}_{t}\in\text{state }i).
\]
We thus briefly call these densities ``landing densities''.
\end{itemize}
Schematically, Deep generative MSMs treat LP1-3 in the way:
\[
\begin{array}{ccc}
\mathbf{x}_{t} & \overset{E}{\rightarrow} & \mathbf{y}_{t}\\
 & \underset{q}{\swarrow}\\
\mathbf{x}_{t+\tau}
\end{array}
\]
Deep generative MSMs represent the transition density (\ref{eq:transition_density})
in the following form (Fig. \ref{fig:network_structures}e):
\begin{equation}
p_{\tau}(\mathbf{x}_{t+\tau}|\mathbf{x}_{t})=E(\mathbf{x}_{t+\tau})^{\top}\mathbf{q}(\mathbf{y};\tau)=\sum_{i=1}^{m}E_{i}(\mathbf{x}_{t})q_{i}(\mathbf{x}_{t+\tau};\tau).\label{eq:DeepGenMSM_transition_density}
\end{equation}
To work with this approach we finally need a Generator $G$, which
is a structure that samples from the density $\mathbf{q}$:
\begin{equation}
G(i,\boldsymbol{\epsilon};\tau)=\mathbf{y}\sim q_{i}(\mathbf{y};\tau)\label{eq:DeepGenMSM_generator}
\end{equation}
It appears that Deep generative MSMs do not learn the propagator explicitly.
However, the propagator can be obtained from $E$ and $\mathbf{q}$
by means the ``rewiring'' trick (Fig. \ref{fig:network_structures}f):
By exchanging the order in which $E$ and $G$ are applied and then
computing the propagator $\mathbf{P}$ as a sample average over $\mathbf{q}$,
obtained from repeatedly applying the generator:
\begin{equation}
p_{ij}(\tau)=\mathbb{E}_{G}\left[E_{j}\left(G(i,\boldsymbol{\epsilon};\tau)\right)\right].\label{eq:DeepGenMSM_Propagator}
\end{equation}
In contrast to VAMPnets (Sec. \ref{sec:VAMPnets}), it is guaranteed
that the propagator (\ref{eq:DeepGenMSM_Propagator}) is a true transition
matrix with nonnegative elements.

\subsection{Deep Resampling MSMs}

We first describe a very simple generator that generates no new (unseen)
configurations, but only learns a function $\mathbf{q}$ that can
be used to resample known configurations \citep{WuEtAl_NIPS18_DeepGenMSM}.
While this approach is clearly limited, it has two advantages: it
will not generate any illegal configuration, and it can be trained
with maximum likelihood. For this approach, we model the landing densities
by

\begin{equation}
q_{i}(\mathbf{x}_{t+\tau})=\frac{w(\mathbf{x}_{t+\tau})\gamma_{i}(\mathbf{x}_{t+\tau})}{\sum_{s=0}^{T-\tau}w(\mathbf{x}_{s+\tau})\gamma_{i}(\mathbf{x}_{s+\tau})}.\label{eq:DeepResampleMSM_landing_density}
\end{equation}
Where $\gamma_{i}(\mathbf{x}_{t+\tau})$ is a trainable, unnormalized
density function and $w$ is an additional weight function which may
be employed to change the weights of configurations, but is usually
identical to 1. In \citep{WuEtAl_NIPS18_DeepGenMSM}, $\gamma_{i}(\mathbf{y})$
is a deep neural network that receives $\mathbf{y}$ as an input as
well as the condition $i$ by means of a one-hot-encoding with $n$
input units, and has a single output node encoding the probability
weight. The normalized density $\mathbf{q}$ is computed by evaluating
the $\gamma$-network for all configurations at time points $\tau,...,T$
and then normalizing over all time points.

Deep resampling MSMs can be trained by maximizing the likelihood based
on expression (\ref{eq:DeepGenMSM_transition_density}), resulting
in the following loss function:
\begin{align*}
\mathcal{L}_{\mathrm{DeepResampleMSM}} & =\sum_{t=1}^{T-\tau}p_{\tau}(\mathbf{x}_{t+\tau}|\mathbf{x}_{t})\\
 & =\sum_{t=1}^{T-\tau}\sum_{i=1}^{m}E_{i}(\mathbf{x}_{t})q_{i}(\mathbf{x}_{t+\tau};\tau)
\end{align*}
where $q_{i}$ is evaluated by (\ref{eq:DeepResampleMSM_landing_density}).
Alternatively, we can optimize $\chi_{i}$ and $\gamma_{i}$ using
the Variational Approach for Markov Processes (VAMP) \citep{WuNoe_VAMP}.
However, we found the ML approach to perform significantly better
in \citep{WuEtAl_NIPS18_DeepGenMSM}. 

In Deep Resample MSMs, the propagator according to (\ref{eq:DeepGenMSM_Propagator})
becomes simply:
\begin{equation}
\mathbf{P}=\frac{1}{T-\tau}\sum_{t=1}^{T-\tau}\mathbf{q}(\mathbf{x}_{t+\tau})E(\mathbf{x}_{t+\tau})^{\top}.
\end{equation}

\begin{figure}
\begin{centering}
\includegraphics[width=1\columnwidth]{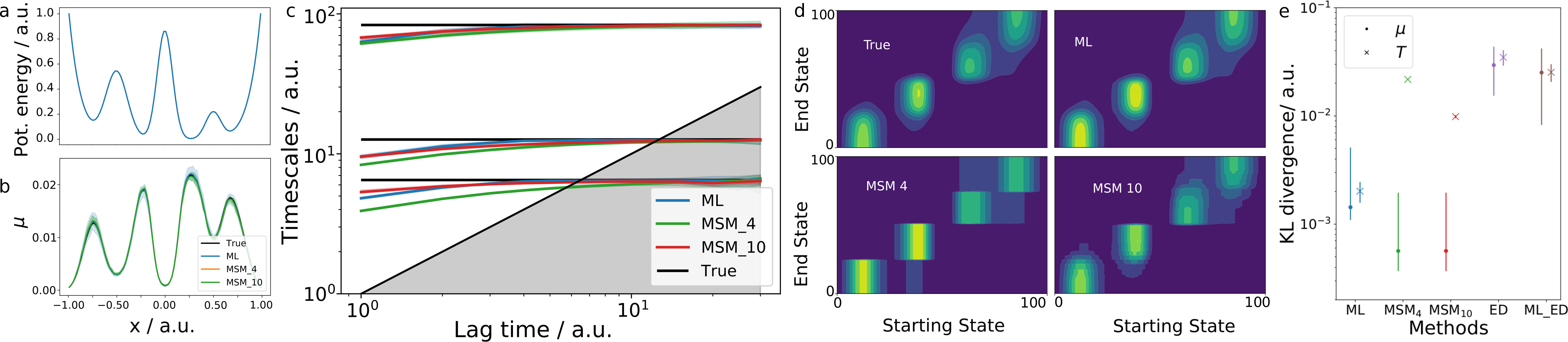}
\par\end{centering}
\caption{\label{fig:DeepMSM_PrinzKinetics}Reproduced from \citep{WuEtAl_NIPS18_DeepGenMSM}:
Performance of deep versus standard MSMs for diffusion in the Prinz
Potential. (a) Potential energy as a function of position $x$. (b)
Stationary distribution estimates of all methods with the exact distribution
(black). (c) Implied timescales of the Prinz potential compared to
the real ones (black line). (d) True transition density and approximations
using maximum likelihood (ML) DeepResampleMSM, four and ten state
MSMs. (e) KL-divergence of the stationary and transition distributions
with respect to the true ones for all presented methods (also DeepResampleMSM).}
\end{figure}

Deep Resample MSMs were found to accurately reproduce the eigenfunctions
and dominant relaxation timescales of benchmark examples \citep{WuEtAl_NIPS18_DeepGenMSM},
and learn to represent the transition density in configuration space
(Fig. \ref{fig:DeepMSM_PrinzKinetics}).

\subsection{Deep Generative MSMs with Energy Distance Loss}

In contrast to resampling MSMs, we now want to generative MSMs, which
can produce genuinely new configurations. This makes the method promising
for performing active learning in MD \citep{BowmanEnsignPande_JCTC2010_AdaptiveSampling,PlattnerEtAl_NatChem17_BarBar},
and to predict the future evolution of the system in other contexts.
To this end, we train a directed generative network to represent (\ref{eq:DeepGenMSM_generator}).
Such a generator can be trained with various principles, e.g. by means
of a variational autoencoder or with adversarial training \citep{KingmaWelling_ICLR14_VAE,GoodfellowEtAl_GANs}.
In \citep{WuEtAl_NIPS18_DeepGenMSM}, we found that a third principle
works well: training the generator $G$ by minimizing the conditional
Energy Distance (ED). The standard ED, introduced in \citep{SzekelyRizzo_Interstat04_EnergyDistance},
is a metric between the distributions of random vectors, defined as
\begin{equation}
D_{E}\left(\mathbb{P}(\mathbf{x}),\mathbb{P}(\mathbf{y})\right)=\mathbb{E}\left[2\left\Vert \mathbf{x}-\mathbf{y}\right\Vert -\left\Vert \mathbf{x}-\mathbf{x}^{\prime}\right\Vert -\left\Vert \mathbf{y}-\mathbf{y}^{\prime}\right\Vert \right]
\end{equation}
for two real-valued random vectors $\mathbf{x}$ and $\mathbf{y}$.
$\mathbf{x}^{\prime},\mathbf{y}^{\prime}$ are independently distributed
according to the distributions of $\mathbf{x},\mathbf{y}$. Based
on this metric, we introduce the conditional energy distance between
the transition density of the system and that of the generative model:
\begin{eqnarray}
D & \triangleq & \mathbb{E}\left[D_{E}\left(\mathbb{P}(\mathbf{x}_{t+\tau}\mid\mathbf{x}_{t}),\mathbb{P}(\hat{\mathbf{x}}_{t+\tau}\mid\mathbf{x}_{t})\right)\mid\mathbf{x}_{t}\right]\nonumber \\
 & = & \mathbb{E}\left[2\left\Vert \hat{\mathbf{x}}_{t+\tau}-\mathbf{x}_{t+\tau}\right\Vert -\left\Vert \hat{\mathbf{x}}_{t+\tau}-\hat{\mathbf{x}}_{t+\tau}^{\prime}\right\Vert -\left\Vert \mathbf{x}_{t+\tau}-\mathbf{x}_{t+\tau}^{\prime}\right\Vert \right]\label{eq:conditional-energy-distance}
\end{eqnarray}
Here $\mathbf{x}_{t+\tau}$ and $\mathbf{x}_{t+\tau}^{\prime}$ are
distributed according to the transition density for given $\mathbf{x}_{t}$
and $\hat{\mathbf{x}}_{t+\tau},\hat{\mathbf{x}}_{t+\tau}^{\prime}$
are independent outputs of the generative model. Implementing the
expectation value with an empirical average results in an estimate
for $D$ that is unbiased, up to an additive constant. We train $G$
to minimize $D$, and subsequently estimate $\mathbf{P}$ by using
the rewiring trick and sampling (\ref{eq:DeepGenMSM_Propagator}).

\begin{figure}
\begin{centering}
\includegraphics[width=0.85\textwidth]{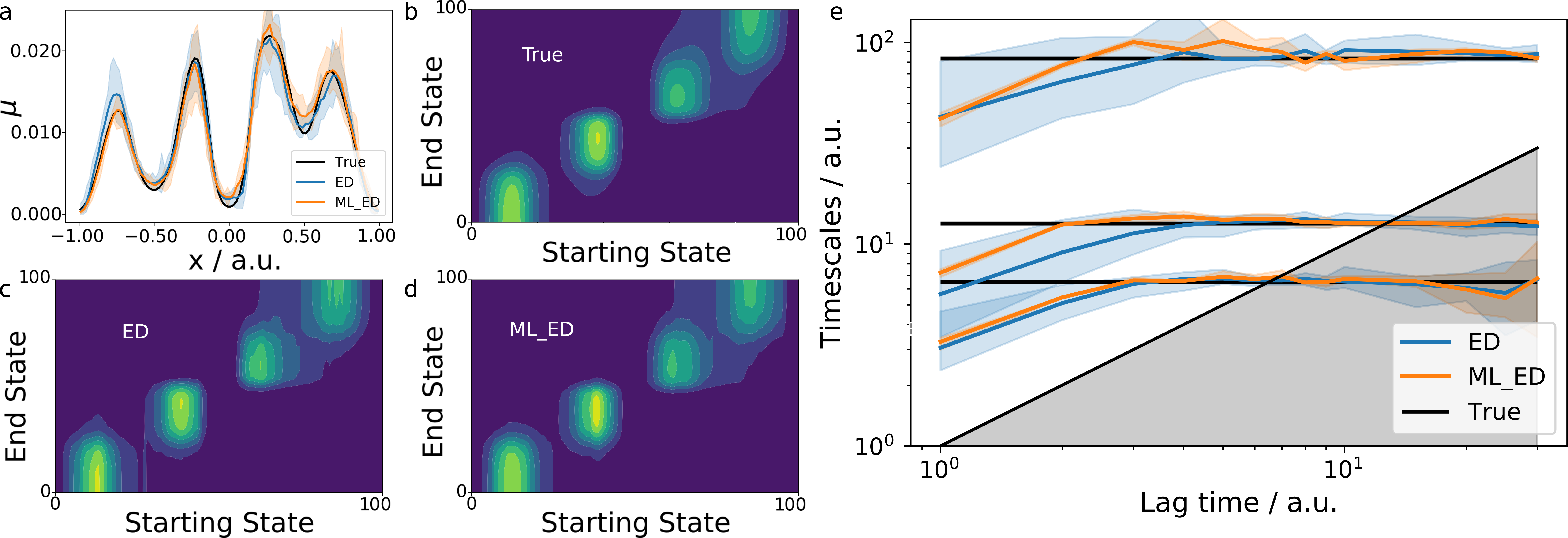}
\par\end{centering}
\caption{\label{fig:PrinzGenerative}Reproduced from \citep{WuEtAl_NIPS18_DeepGenMSM}:
Performance of deep generative MSMs for diffusion in the Prinz Potential.
Comparison between exact reference (black), deep generative MSMs estimated
using only energy distance (ED) or combined ML-ED training. (a) Stationary
distribution. (b-d) Transition densities. (e) Relaxation timescales.}
\end{figure}

\begin{figure}
\begin{centering}
\includegraphics[width=1\columnwidth]{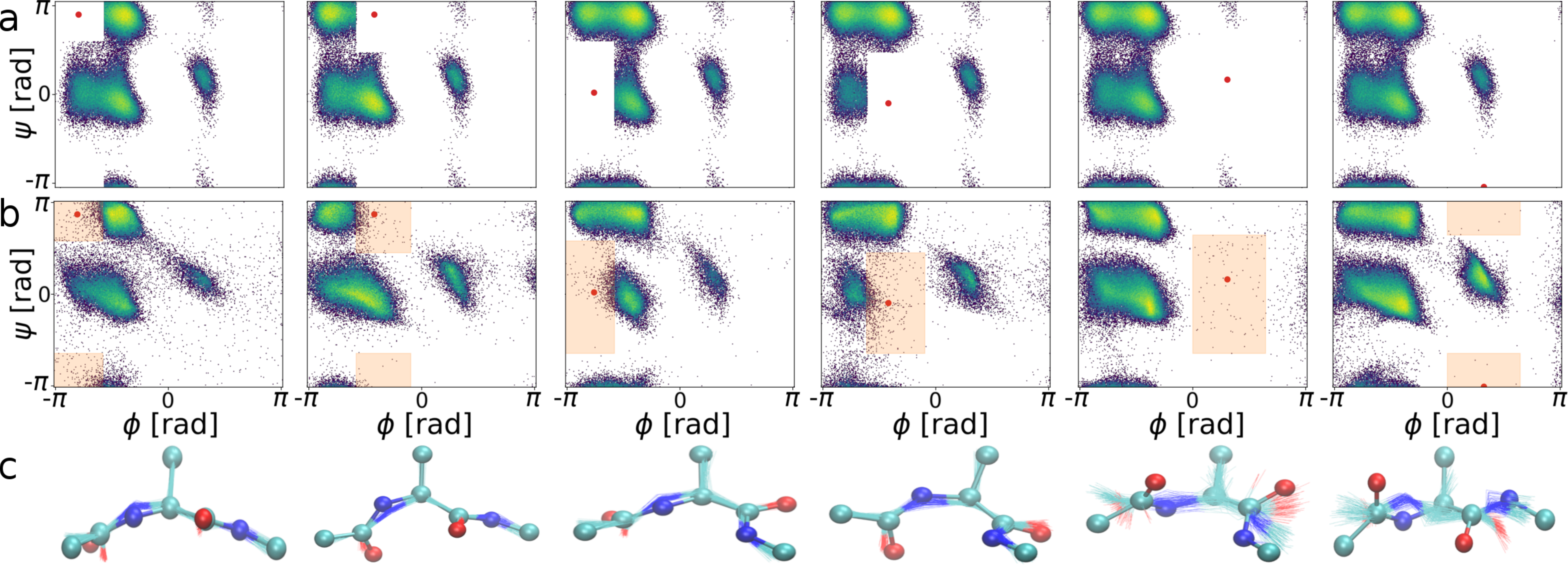}
\par\end{centering}
\caption{\foreignlanguage{english}{\label{fig:AlaGenerative}\foreignlanguage{american}{Reproduced from
\citep{WuEtAl_NIPS18_DeepGenMSM}: DeepGenMSMs can generate physically
realistic structures in areas that were not included in the training
data. (a) Distribution of training data. (b) Generated stationary
distribution. (c) Representative ``real'' molecular configuration
(from MD simulation) in each of the metastable states (sticks and
balls), and the 100 closest configurations generated by the deep generative
MSM (lines).}}}
\end{figure}

Deep Generative MSMs trained with the energy distance were also found
to accurately reproduce the eigenfunctions and dominant relaxation
timescales of benchmark examples \citep{WuEtAl_NIPS18_DeepGenMSM},
and learn to represent the transition density in configuration space
(Fig. \ref{fig:PrinzGenerative}). In contrast to Resampling MSMs
described in the previous section, they can also be used to generalize
to sampling new, previously unseen, configurations, and are therefore
a first approach to sample the long-time propagator $\mathbf{x}_{t}\rightarrow\mathbf{x}_{t+\tau}$
in configuration space (Fig. \ref{fig:AlaGenerative}).


\section{Data and Software}

Many of the algorithms described above are implemented in the PyEMMA
\citep{SchererEtAl_JCTC15_EMMA2,SenneSchuetteNoe_JCTC12_EMMA1.2}
software -- \href{http://www.pyemma.org}{www.pyemma.org} and in
MSMbuilder \citep{HarriganEtAl_BJ17_MSMbuilder}. Some of the deep
learning algorithms can be found at \href{https://github.com/markovmodel/deeptime}{https://github.com/markovmodel/deeptime}. 

The field is still lacking good resources with public datasets, partially
because long-time MD data of nontrivial systems is typically extremely
large (giga- to terabytes), and due to the unsupervised nature of
the learning problems, the role of a benchmarking dataset is less
straightforward as in supervised learning. Commonly used datasets
for the evaluation of long-time MD models are the fast folding protein
trajectories produced by D. E. Shaw research on the Anton supercomputer
\citep{LindorffLarsenEtAl_Science11_AntonFolding}, which can be obtained
from them on request. We provide datasets for small peptides via the
Python package mdshare \href{https://markovmodel.github.io/mdshare/}{https://markovmodel.github.io/mdshare/}.

\end{document}